\documentclass{IEEEtran}
\usepackage{mathrsfs}
\usepackage{pifont}
\usepackage{bbding}
\usepackage{tipa}
\usepackage{color}
\usepackage{cite}
\usepackage{epsfig}
\usepackage{graphicx}
\usepackage{url}
\usepackage{amsfonts}
\usepackage{multicol}
\usepackage{float}
\usepackage{times}
\usepackage{psfrag}
\usepackage{subfigure}
\usepackage{stfloats}
\usepackage{amsfonts,amsmath,amssymb}
\usepackage{footnote}
\usepackage{array}
\usepackage{epic}
\usepackage{graphicx}
\usepackage{curves}
\usepackage{balance}
\usepackage{algorithm}
\usepackage{algorithmic}
\usepackage{multirow}
\usepackage{xcolor}
\usepackage{graphics}
\usepackage{epstopdf}
\usepackage{enumerate}
\usepackage{geometry}
\begin{document}

\title{Probabilistic Caching for Small-Cell Networks with Terrestrial and Aerial Users}
\author{Fei Song, Jun Li,~\IEEEmembership{Senior Member,~IEEE,} Ming Ding,~\IEEEmembership{Senior Member,~IEEE,}\\
Long Shi,~\IEEEmembership{Member,~IEEE,} Feng Shu,~\IEEEmembership{Member,~IEEE,} Meixia Tao,~\IEEEmembership{Fellow,~IEEE,} \\
Wen Chen,~\IEEEmembership{Senior Member,~IEEE,} H. Vincent Poor,~\IEEEmembership{Fellow,~IEEE}


 \thanks{Copyright (c) 2015 IEEE. Personal use of this material is permitted. However, permission to use this material for any other purposes must be obtained from the IEEE by sending a request to pubs-permissions@ieee.org.}
 \thanks{Manuscript received xxx xxx, xxx; revised xxx xxx, xxx; accepted xxx xxx, xxx.
 This work was supported in part
 by the National Key R\&D Program under Grant 2018YFB1004800,
 by the National Natural Science Foundation of China under Grants 61872184, 61727802, 61571299 and 61671294,
 by the STCSM Key Fundamental Project under Grants 16JC1402900 and 17510740700,
 by the National Science and Technology Major Project under Grant 2018ZX03001009-002,
 by the U.S. National Science Foundation under Grants CCF-0939370 and CCF-1513915. \emph{(Corresponding authors: Long Shi, Jun Li.)}  }
 \thanks{F. Song, J. Li and F. Shu are with the School of Electronic and Optical Engineering, Nanjing University of Science Technology, Nanjing 210094, China (e-mail: \{fei.song, jun.li, shufeng\}@njust.edu.cn).}
 \thanks{M. Ding is with the Data61, CSIRO, Sydney, N.S.W. 2015, Australia (e-mail: Ming.Ding@data61.csiro.au).}
 \thanks{L. Shi is with the Science and Math Cluster, Singapore University of Technology and Design, Singapore 487372 (e-mail: slong1007@gmail.com).}
 \thanks{M. Tao and W. Chen are with Shanghai Institute of Advanced Communications and Data Sciences, Department of Electronic Engineering, Shanghai Jiao Tong University, Shanghai 200240, China (e-mail: \{mxtao, wenchen\}@sjtu.edu.cn).}
 \thanks{H. V. Poor is with Department of Electrical Engineering, Princeton University, NJ 08544, USA (e-mail: Poor@princeton.edu.)}
 }


\maketitle

\vspace{-20mm}
\begin{abstract}
The support for aerial users has become the focus of recent 3GPP standardizations of 5G, due to their high maneuverability and flexibility for on-demand deployment. In this paper, probabilistic caching is studied for ultra-dense small-cell networks with terrestrial and aerial users, where a dynamic on-off architecture is adopted under a sophisticated path loss model incorporating both line-of-sight and non-line-of-sight transmissions. Generally, this paper focuses on the successful download probability (SDP) of user equipments (UEs) from small-cell base stations (SBSs) that cache the requested files under various caching strategies. To be more specific, the SDP is first analyzed using stochastic geometry theory, by considering the distribution of such two-tier UEs and SBSs as Homogeneous Poisson Point Processes. Second, an optimized caching strategy (OCS) is proposed to maximize the average SDP. Third, the performance limits of the average SDP are developed for the popular caching strategy (PCS) and the uniform caching strategy (UCS). Finally, the impacts of the key parameters, such as the SBS density, the cache size, the exponent of Zipf distribution and the height of aerial user, are investigated on the average SDP. The analytical results indicate that the UCS outperforms the PCS if the SBSs are sufficiently dense, while the PCS is better than the UCS if the exponent of Zipf distribution is large enough. Furthermore, the proposed OCS is superior to both the UCS and PCS.

\end{abstract}


  \section{Introduction}
  With the dramatic proliferation of smart mobile devices and various mobile applications, the global mobile data traffic has been increasing rapidly in recent years. Cisco forecasts that the traffic will increase sevenfold from 2016 to 2021, of which about 78 percent will be video streams by 2021~\cite{Cisco1}. This deluge of data has driven vendors and operators to seek every possible tool at hand to improve network capacity \cite{UD_SC_Deployments}. Very recently, providing wireless connectivity for unmanned aerial vehicles (UAVs) has become an emerging research area~\cite{LTE_for_UAV}. The UAV becomes increasingly popular for various commercial, industrial, and public-safety applications \cite{AD1}.
  Due to their high maneuverability and flexibility for on-demand deployment, UAVs equipped with advanced transceivers and batteries are gaining increasing popularity in information technology applications \cite{Trajectory_Communication}, and have been widely used in delivery, communications and surveillance.
  As the UAV applications proliferate, security issues in the UAV deployment have captured much attention in recent years.
  Hence, the academia and industry have expanded public safety communications from the ground \cite{AD2} to the air \cite{AD3},~\cite{AD4}.
  Driven by the rising interest in aerial communications, the Third Generation Partnership Project (3GPP) has taken UAVs supported by Long Term Evolution (LTE) as a primary research focus \cite{RP_170779}.
  \vspace{-2mm}
  \subsection{Background of Wireless Caching}
  Recent research has unveiled that some popular files are repeatedly requested by the user equipments (UEs), which takes a huge portion of the data traffic \cite{Cache_Not}. To reduce duplicated transmissions, wireless caching has been proposed to pre-download the popular files in cache devices at the wireless edges \cite{D2D_Assisted_Caching}, \cite{D2D_caching}.
  Unlike the traditional communication resources, the storage resources are abundant, economical, and sustainable, making the caching technology even more promising in the modern communications~\cite{Caching_NTier}. For example, a scalable platform for implementing the caching technology is well-known as the small-cell caching in the ultra-dense (UD) small-cell networks (SCNs), which has attracted significant attention as one of the enticing approaches to meet the ever-increasing data traffic demands for the fifth generation (5G) communication systems \cite{UD_SC_Deployments}, \cite{Contract}. In the small-cell caching, popular files are pre-downloaded into local caches of small-cell base stations (SBSs) in the off-peak hours, and ready to be fetched by the UEs in the peak hours, alleviating the backhaul congestion in wireless networks.
  In addition, small-cell caching makes the data traffic much closer to the mobile users. Thus, the transmission latency can be reduced, and the quality of experience for users will be enhanced.
  However, each device has a finite amount of storage, popular content should be seeded into the network in a way that maximizes the successful download probability (SDP).
  \vspace{-2mm}
  \subsection{Related Work}
  Existing works have shown that the file placement of small-cell caching largely follows two approaches: deterministic placement and probabilistic placement. For deterministic placement, files are placed and optimized for specific networks by a deterministic pattern \cite{Distributed_Caching,Mobile_Data_Caching,Pricing}. In practice, the wireless channels and the geographic distribution of UEs are time-variant.
  This triggers the optimal file placement strategy to be frequently updated, which makes the file placement highly complicated.
  To cope with this problem, probabilistic file placement considers that each SBS randomly caches a subset of popular files with a certain caching probability in the stochastic networks. As a seminal work, \cite{Cache_enabled_SCN} modelled the node locations as Homogeneous Poisson Point Processes (HPPPs) and analyzed the general performance of the small-cell caching. Compared with caching the same copy of certain files in all SBSs, probabilistic file placement in the small-cell caching is more flexible and robust.

  However, the cache-aided ground SBSs may not be able to support the users in high rise building scenarios \cite{caching_sky} and in emergency situations where the ground infrastructures fail or there is a sudden and temporary surge of traffic demand~\cite{LTE_sky}.
  The UAVs with high maneuverability and flexibility can be used as flying relays~\cite{UAV_Suburban} to dynamically cache the popular content files from the cache-aided ground BSs and then effectively disseminate them to the users \cite{Learn_Cache}.
  Zhao $et$ $al$.~\cite{Caching_UAV} studied the cache-enabled UAVs that serve as flying BSs and refresh the cached content files from macro BSs (MBSs).
  However, it is time and energy consuming for UAVs with limited battery capacity to fly back to MBSs to update the cached content files. In view of this problem, the UD SCNs point out a promising UAV-aided wireless caching scenario where UAVs are connected to the nearby SBSs that are much denser than MBSs. In this scenario, the role of UAV can be either terminal UE served by static BSs or flying relays that forward files to other UEs, aiming at alleviating the peak backhaul traffic and assisting the SBSs.
  Recent works \cite{LTE_sky}, \cite{UAV_Suburban} and \cite{UAV_MIMO} have extended conventional terrestrial cellular services to aerial users in the 5G networks. In addition, the 3GPP launched an investigation on enhanced LTE support for aerial users in 2017 and proposed a channel model for aerial UEs~\cite{TR_36777}. Therefore, in this paper, we focus on the small-cell cellular networks with terrestrial users (TUs) and aerial users (AUs), i.e., UAVs.

  From the stochastic cellular network model, the BS locations are supposed to follow an HPPP distribution \cite{Tractable_Approach}, \cite{KTier_Downlink}. Ref. \cite{Optimal_geographic_caching} proposed an optimal geographic placement in wireless cellular networks modelled by HPPP. Furthermore, a trade-off between the SBS density and the storage size was presented in \cite{Cache_enabled_SCN}, where each SBS caches the most popular files.
  In \cite{Probabilistic_SC_Caching}, the library is divided into $N$ file groups and the probabilistic caching probability of each file group is optimized to maximize the SDP in SCNs. Utilizing stochastic geometry, \cite{Caching_Multicasting} optimized probabilistic caching at helper stations in a two-tier heterogeneous network, where one tier of multi-antenna MBSs coexists with the other tier of helpers with caches. However, to our best knowledge, most existing works on small-cell caching considered the path loss models without differentiating line-of-sight (LoS) from non-line-of-sight (NLoS) transmissions. It is well known that LoS transmission may occur when the distance between a transmitter and a receiver is small and no shelter, and NLoS transmission is common in indoor environments and in central business districts. In our previous works \cite{LoS_NLoS},~\cite{Small_Cell_Caching}, we considered both multi-slope piece-wise path loss function and probabilistic LoS or NLoS transmission in cellular networks. Furthermore, for ease of exposition, we ignored the antenna height difference between SBSs and UEs in the performance analysis due to the dominance of the horizontal distance. However, the antenna height difference becomes non-negligible as the distance between an UE and its serving SBS decreases. To verify this, \cite{Antenna_Heights} clarified that the height difference between UEs and BSs imposes a significant impact on coverage probability and area spectral efficiency.

  Regarding the SBS activity, there are two network architectures in the SCNs, namely, the always-on architecture and the dynamic on-off architecture. The always-on architecture is commonly used in the current cellular networks, where all the SBSs are always active. By contrast, in the dynamic on-off architecture, the SBSs are only active when they are required to provide services to UEs~\cite{Femtocell_BS}. To mitigate inter-cell interference, the dynamic on-off architecture will thrive as an important 5G technology in the UD SCNs, which is also investigated in 3GPP \cite{UD_SC_Deployments}. Therefore, in this paper, we focus on the dynamic on-off architecture.
  \vspace{-3mm}
  \subsection{Contributions}
  In this work, we analyze the average SDP that UEs can successfully download files from the storage of SBSs and optimize the caching probability of each file.
  We consider an UD SCN with UEs including TUs and AUs under a general path loss model that incorporates both LoS and NLoS paths. Furthermore, we consider the dynamic on-off architecture.
  Our goal is to maximize the average SDP. To be concise, the contributions of this article are summarized as follows:
  \begin{itemize}
    \item We investigate the average SDP by considering the 3GPP path loss models for TUs and UAVs respectively (see Section IV). 
    \item We propose the optimized caching strategy (OCS) to maximize the average SDP of UD SCNs by optimizing caching probability of each content (see Section V).
    \item We analyze the performance limits of the SDP with the uniform caching strategy (UCS) and the popular caching strategy (PCS) under a single-slope path loss model, respectively (see Section VI). First, we show that the OCS is superior to both the UCS and PCS. Second, we reveal that the UCS outperforms the PCS if the SBS density is large enough, while the PCS is better than the UCS if the exponent of Zipf distribution grows sufficiently large.
  \end{itemize}

  The rest of the paper is organized as follows. We describe the system model in Section II and study the probabilistic caching strategy in Section III. Section IV presents our analytical results of SDP. Section V proposes the OCS. Section VI shows the impacts of network parameters under the UCS and PCS. Section VII provides simulations and numerical results. Finally, Section VIII concludes this paper. Table~I lists main notations and symbols used in this paper.

  \section{System Model}
  As illustrated in Fig. 1, we consider a small-cell cellular network where the SBSs serve two-tier UEs including TUs as ground users and UAVs as AUs over the same frequency spectrum. We assume that the SBSs, the TUs and the UAVs are deployed according to three independent HPPPs with the height of $h_{\rm{BS}}$, $h_{\rm{TU}}$ and $h_{\rm{AU}}$, respectively.
  With reference to \cite{Trajectory_Communication} and \cite{UAV_MIMO}, consider that all the UAVs are positioned at the same height\footnote[1]{This paper considers that the locations of UAVs follow a 2D HPPP with the same height. As Fig. 7 will show, the change of the UAV height affects its average SDP but imposes no performance impact on TUs. We remark that the optimization of the caching probabilities (to be discussed in Section V) when the UAVs are deployed as a 3D HPPP is quite challenging and will be left as our future work.}. Let $\lambda_{\rm{s}}$ be the density of SBSs, $\lambda_{\rm{TU}}$ be the density of TUs, and $\lambda_{\rm{AU}}$ be the density of UAVs. Second, each UE\footnote[2]{By the UE, we mean either the TU or the UAV.} is associated with an SBS with the smallest path loss. The transmit power of each SBS is denoted by $P$.

\begin{table}[t]
 \newcommand{\tabincell}[2]{\begin{tabular}{@{}#1@{}}#2\end{tabular}}
  \centering
\caption{Main notations and their definitions}
\begin{tabular}{|c|c|}
    \hline
    \multicolumn {1}{|c|}{Notations} & \multicolumn {1}{c|}{Definitions}                            \\
    \hline
    $h_{\rm{BS}},h_{\rm{TU}},h_{\rm{AU}}$     &Height of SBSs, TUs, and AUs                           \\
    \hline
    $\lambda_{\rm{s}},\lambda_{\rm{u}},\lambda_{\rm{TU}},\lambda_{\rm{AU}}$      & Density of SBSs, UEs, TUs, and AUs       \\
    \hline
    $P$  &Transmit power of each SBS                                                                 \\
    \hline
    $h,r,l$  &\tabincell{c}{Absolute antenna height difference, horizontal \\distance, and distance between SBS and UE}           \\
    \hline
    $l_{\rm{TU}},l_{\rm{AU}}$  &\tabincell{c}{Distance between TU and SBS, \\ distance between AU and SBS}     \\
    \hline
    $Q_n, S_n$       &\tabincell{c}{Request probability of the $n$-th file, \\ caching probability of the $n$-th file}   \\
    \hline
    $D_n$       &\tabincell{c}{Event that the typical UE can successfully \\receive its requested $n$-th file }   \\
    \hline
    $A_n$       &\tabincell{c}{Event that the SBS in the $n$-th tier is active }   \\
    \hline
     $\rm{Pr}(D_n)$  &Successful download probability \\
    \hline
    ${\overline{\Pr}}$  &Average successful download probability \\
    \hline
    $M,N,\beta$    &\tabincell{c}{Number of files, cache size,\\ and exponent of Zipf distribution}  \\
    \hline
    ${\rm{Pr}^L}_k(\cdot)$   &\tabincell{c}{$k$-th piece LoS probability function}  \\
    \hline
    $\zeta_k(\cdot)$   &\tabincell{c}{$k$-th piece path loss function}  \\
    \hline
    $f_k^{\rm{L}}(\cdot),f_k^{\rm{NL}}(\cdot)$  &PDFs for LoS path and NLoS path \\
    \hline
  \end{tabular}
  \end{table}

  The horizontal distance between an SBS and an UE is denoted by $r$. Moreover, the absolute antenna height difference between an UE and an SBS is denoted by $h$, and the distance between an UE and an SBS is denoted by $l$.
  Let $l_{\rm{TU}}$ be the distance between a TU and an SBS, and $l_{\rm{AU}}$ be the distance between an AU and an SBS. As such, the height differences are $h_1=h_{\rm{BS}}-h_{\rm{TU}}$ for TUs and $h_2=h_{\rm{AU}}-h_{\rm{BS}}$ for UAVs respectively. Hence, the distance $l$ can be expressed as
   \begin{equation}
  l = \sqrt {{r^2} + {h^2}} .
  \end{equation}

  Regarding the UAV acting as the aerial UE, recent works such as \cite{LTE_for_UAV}, \cite{LTE_sky}, \cite{UAV_Suburban} and \cite{UAV_MIMO} have studied a variety of communication scenarios where conventional terrestrial cellular services are extended to aerial UEs. 

  Considering the downlink transmission, the small-scale effect in the network is assumed to be Rayleigh fading, and the path loss model embraces both LoS and NLoS paths as large-scale fading. The link from any UE to the typical SBS has a LoS path with probability ${\Pr ^{\rm{L}}}(r,h)$ or a NLoS path with probability $1 - {\Pr ^{\rm{L}}}(r,h)$, respectively. According to \cite{TR_36777} and \cite{TR_36828}, the piece-wise LoS probability function is given by
  \begin{equation}
  {\rm{Pr}^{{\rm{L}}}}\left( r,h \right) = \left\{ {\begin{array}{*{20}{l}}
{\Pr _1^{\rm{L}}\left( r,h \right),}\\
{\Pr _2^{\rm{L}}\left( r,h \right),}\\
 \vdots \\
{\Pr _K^{\rm{L}}\left( r,h \right),}
\end{array}} \right.\begin{array}{*{20}{l}}
{0 < r \le {d_1(h)}}\\
{{d_1(h)} < r \le {d_2(h)}}\\
 \vdots \\
{r > {d_{K - 1}(h)}}
\end{array},
  \end{equation}
  where $\Pr _k^{\rm{L}}\left( r,h \right)$, $k \in \{1, 2, ... , 	K\}$ is the $k$-th piece probability function that an UE and an SBS separated by a horizontal distance ${d_{k-1}(h)} < r \le {d_k}(h)$ has a LoS path. In addition, ${d_k}(h)$ varies as the height difference $h$ changes. The details of ${d_k}(h)$ for TUs and UAVs are provided in Sections IV-C and IV-D, respectively.

  \begin{small}
  \begin{figure*}[ht]
  \begin{equation}
  \label{equ:1}
  \zeta \left( r,h \right) = \left\{ {\begin{array}{*{20}{l}}
{\zeta_1\left( r,h \right) = \left\{{\begin{array}{*{20}{l}}
{A_1^{\rm{L}}{l^{{\rm{ - \alpha }}_{\rm{1}}^{\rm{L}}}},\;{\text{LoS with}}\;\Pr _1^{\rm{L}}\left( r,h \right)}\\
{A_1^{{\rm{NL}}}{l^{{\rm{ - \alpha }}_{\rm{1}}^{{\rm{NL}}}}},\;{\text{NLoS with}}\;\left( {1 - \Pr _1^{\rm{L}}\left( r,h \right)} \right)}
\end{array},} \right.}&{0 < r \le {d_1(h)}}\\
{\zeta_2\left( r,h \right) =\left\{ {\begin{array}{*{20}{l}}
{A_2^{\rm{L}}{l^{{\rm{ - \alpha }}_{\rm{2}}^{\rm{L}}}},\;{\text{LoS with}}\;\Pr _2^{\rm{L}}\left( r,h \right)}\\
{A_2^{{\rm{NL}}}{l^{{\rm{ - \alpha }}_{\rm{2}}^{{\rm{NL}}}}},\;{\text{NLoS with}}\;\left( {1 - \Pr _2^{\rm{L}}\left( r,h \right)} \right)}
\end{array},} \right.}&{{d_1(h)} < r \le {d_2(h)}}\\
 \vdots & \vdots \\
{\zeta_K\left( r,h \right) =\left\{ {\begin{array}{*{20}{l}}
{A_K^{\rm{L}}{l^{{\rm{ - \alpha }}_K^{\rm{L}}}},\;{\text{LoS with}}\;\Pr _K^{\rm{L}}\left( r,h \right)}\\
{A_K^{{\rm{NL}}}{l^{{\rm{ - \alpha }}_K^{{\rm{NL}}}}},\;{\text{NLoS with}}\;\left( {1 - \Pr _K^{\rm{L}}\left( r,h \right)} \right)}
\end{array},} \right.}&{r > {d_{K - 1}(h)}}
\end{array}} \right..
  \end{equation}
  \hrulefill
  \end{figure*}
  \end{small}
  Furthermore, with reference to \cite{TR_36777} and \cite{TR_36828}, we adopt a general and practical path loss model in \cite{LoS_NLoS}, where the path loss with respect to the distance $l$ is modeled as (\ref{equ:1}) (see the top of next page).
  In (\ref{equ:1}), $A_k^{\rm{L}}$ and $A_k^{\rm{NL}}$ denote the path losses of the LoS path and NLoS path at a reference distance of $l = 1$ respectively. Moreover, $\alpha_k^{\rm{L}}$ and $\alpha_k^{\rm{NL}}$ denote the path loss exponents of the LoS path and NLoS path respectively. It is worthwhile noting that the values of $A_k^{\rm{L}}$, $A_k^{\rm{NL}}$, $\alpha_k^{\rm{L}}$ and $\alpha_k^{\rm{NL}}$ for the TUs are different from those for the UAVs. For TUs, $A_k^{\rm{L}}$, $A_k^{\rm{NL}}$, $\alpha_k^{\rm{L}}$ and $\alpha_k^{\rm{NL}}$ are constants from field tests in~\cite{TR_36828}. For UAVs, $A_k^{\rm{L}}$ and $A_k^{\rm{NL}}$ are constants, while $\alpha_k^{\rm{L}}$ and $\alpha_k^{\rm{NL}}$ vary with different height ranges \cite{TR_36777}.

  In the dynamic on-off architecture, an SBS is only active when it is required to serve the associated UEs. At any time, a typical UE only associates with an intended SBS, while other active SBSs are regarded as interferers.
  Since both TU and UAV can be regarded as the typical UE in this model, the two-tier UEs need to be projected onto the same plane for this architecture.
  As such, we can get the total density of two-tier UEs ${\lambda _{\rm{u}}} = {\lambda _{\rm{TU}}} + {\lambda _{\rm{AU}}}$. Hence, the probability that an SBS is active is given by \cite{Idle_Mode}
  \begin{equation}
  \label{equ:pr_on}
  {\rm{Pr}_{{\rm{on}}}} \approx 1 - {\left( {1 + {\textstyle{{{\lambda_{\rm{u}}}} \over {{q }{\lambda_{\rm{s}}}}}}} \right)^{ - {q }}},
  \end{equation}
 where $q=$ 3.5 is a tight lower bound of $q$, especially for dense SCNs.

\begin{figure}[t]
 \centering
    \includegraphics[width=2.5in,angle=0]{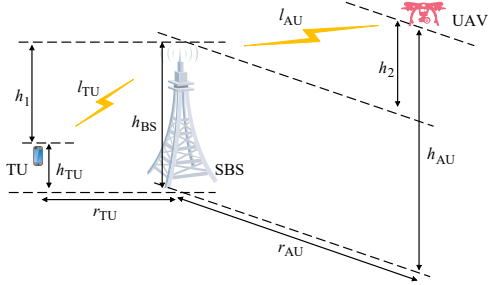}
    \caption{System model of a small-cell network consisting of the SBS and two-tier UEs.}
\end{figure}

\section{Probabilistic UD SCNs Caching Strategy}

  Suppose that a library consists of $N$ popular files each with equal length. Note that $N$ represents the number of popular files that the UEs tend to access rather than the number of files available on the Internet.
  Furthermore, each file is requested according to its popularity, known as $a$ $priori$ information.

  Let $Q_n$ represents the probability that the $n$-th file is requested by the UEs. ${\bf{Q}} = \left[ {Q_1}, {Q_2}, \cdots, {Q_N} \right]$ collects the request probability mass functions (PMFs) of all $N$ files.
  Similar to existing works \cite{Caching_NTier}, \cite{Probabilistic_SC_Caching} and \cite{Edge_Caching}, we model the PMF of each file request as Zipf distribution, and the request probability is given by
  \begin{equation}
  \label{equ:22}
  {Q _n} = \frac{{{\textstyle{1 \over {{n^\beta }}}}}}{{\sum\nolimits_{i = 1}^N {{\textstyle{1 \over {{i^\beta }}}}} }},
  \end{equation}
  where $\beta$ is the exponent of the Zipf distribution. A larger $\beta$ implies a more uneven popularity among those files. 

  Due to the limited storage, each SBS cannot cache the entire file library. In this context, we consider that the files are independently placed in different SBSs. Suppose that a cache memory of size $M$ is available on each SBS.
  In the file placement phase, each SBS store the $n$-th file in its local cache with a caching probability $S_n$, yielding
  \begin{align}
  \sum\limits_{n = 1}^N {{S_n} \leq M,}\\
  0 \le {S_n} \le 1,\forall n.
  \end{align}

  We emphasize that the $n$-th tier of SBS is formed by a group of SBSs that cache the $n$-th file. Given that each SBS independently caches the files, the distribution of SBSs that cache the $n$-th file is viewed as a thinned HPPP with density of ${S_n}{\lambda_{\rm{s}}}$. In addition, given that each UE only requests a single content at each time slot, the distribution of UEs who request the $n$-th file can also be modeled as a thinned HPPP with density of ${Q_n}{\lambda_{\rm{u}}}$.
  In the following, we consider three types of caching strategies:
  \begin{enumerate}
    \item UCS: Each SBS caches each file randomly with equal probability \cite{Caching_NTier}.
    \item PCS: Each SBS only caches the most popular files  \cite{Caching_NTier}.
    \item OCS: Each SBS caches each file with optimized probability for SDP maximization (see Section V).
  \end{enumerate}

\section{Performance Analysis of Small-Cell Caching}

  In this section, we derive the SDP for the dynamic on-off architecture. Some cases adopted by the 3GPP are also considered.
\vspace{-3mm}
\subsection{Received SINR}
  The received signal power of a typical UE from its associated SBS can be written as
  \begin{equation}
  {P_{{\rm{rs}}}} = P\zeta \left( r,h \right)g,
  \end{equation}
  where the channel gain of the Rayleigh fading $g$ follows an independent and identically distributed ($i.i.d.$) exponential distribution with unit mean.

  Consequently, the signal-to-interference-and-noise-ratio (SINR) at the typical UE can be expressed as
  \begin{equation}
  {\rm{SINR}} = \frac{{{P_{{\rm{rs}}}}}}{{{\sigma ^{\rm{2}}} + {I_Z}}},
  \end{equation}
  where $\sigma^2$ is noise power, and $Z$ is the set of interfering SBSs with the total interference being
  \begin{equation}
  {I_Z} = \sum\limits_{z \in Z} {P\zeta \left( {{r_z}},h \right)} {g_z},
  \end{equation}
  where $g_z$ denotes the channel gain between the typical user and the $z$-th interfering SBS, also following an $i.i.d.$ exponential distribution with unit mean.
\vspace{-3mm}
\subsection{Successful Download Probability }

  Let $D_n$ be the event that the typical UE can successfully receive the requested $n$-th file from the associated $n$-th tier of SBS. In this paper, we consider that $D_n$ occurs if the SINR of the UE is no less than a targeted value $\delta$. As such, the SDP of $D_n$ can be formulated as
  \begin{equation}
  \Pr \left( {{D_n}} \right) = \Pr \left( {{\rm{SINR}} > \delta } \right).
  \end{equation}

  Recall that the SBSs in the $n$-th tier and the UEs that request the $n$-th file form two independent thinned HPPPs
  with densities ${S_n}{\lambda_{\rm{s}}}$ and ${Q_n}{\lambda_{\rm{u}}}$ respectively.
  Let $A_n$ be the event that the SBS in the $n$-th tier is active, we rewrite the probability that an SBS in the $n$-th tier is active as
  \begin{equation}
  \label{equ:6}
  {\Pr(A_n)} \approx 1 - {\left( {1 + {\textstyle{{{Q_n}{\lambda _{\rm{u}}}} \over {{q }{S_n}{\lambda _{\rm{s}}}}}}} \right)^{ - {q}}}, 
  \end{equation}
  where we replace $\lambda_{\rm{s}}$ and $\lambda _{\rm{u}}$ in (\ref{equ:pr_on}) with ${S_n}{\lambda _{\rm{s}}}$ and ${Q_n}{\lambda _{\rm{u}}}$ respectively.

  $Theorem$ 1: Given a particular value of ${S_n}{\lambda_{\rm{s}}}$, the SDP of $D_n$ is given by
  \begin{equation}
  \Pr \left( {{D_n}} \right) = \sum\limits_{k = 1}^K {\left( {T_k^{\rm{L}} + T_k^{{\rm{NL}}}} \right)} ,
  \end{equation}
  where
  \begin{align}
  \label{equ:2}
  T_k^{\rm{L}} = \int_{{d_{k - 1}(h)}}^{{d_k}(h)} {\Pr \left[ {\frac{{P\zeta _k^{\rm{L}}\left( r,h \right)g}}{{{\sigma ^{\rm{2}}} + {I_Z}}} > \delta } \right]} f_k^{\rm{L}}(r,h){\rm{d}}r,\\
  \label{equ:3}
  T_k^{{\rm{NL}}}\!=\! \int_{{d_{k - 1}}(h)}^{{d_k}(h)}\!{\Pr \!\left[ {\frac{{P\zeta _k^{{\rm{NL}}}\left( r,h \right)g}}{{{\sigma ^{\rm{2}}} + {I_Z}}}\!>\! \delta } \right]\!} f_k^{{\rm{NL}}}(r,h){\rm{d}}r,
  \end{align}
  and $f_k^{\rm{L}}\left( r,h \right)$ and $f_k^{{\rm{NL}}}\left( r,h \right)$ are the probability density functions (PDFs) of LoS path and NLoS path, respectively. Let $d_0 = 0$ and $d_K = \infty$. Moreover, we have
  \begin{align}
  \label{equ:9}
  f_k^{\rm{L}}\!\left( r,h \right) = &\exp \!\left( \!{ - \!\int_0^{{r_1}} {2\pi{S_n}{\lambda _{\rm{s}}}\!\left(\! {1 \!-\! {\rm{Pr}} _k^{\rm{L}}\left( u,h \right)} \!\right)\!u{\rm{d}}u} } \!\right)\!\nonumber\\
 \times &\exp \left( { - \int_0^r {2\pi {S_n}{\lambda _{\rm{s}}}{\rm{Pr}} _k^{\rm{L}}\left( u,h \right)u{\rm{d}}u} } \!\right)\! \nonumber\\
\times & {\rm{Pr}}_k^{\rm{L}}\!\left( r,h \right)\! 2\pi r{S_n}{\lambda _{\rm{s}}},\; {{d_{k\! -\! 1}(h)} \!<\! r \!\le\! {d_k}(h)},
  \end{align}
  \begin{align}
  \label{equ:10}
  f_k^{{\rm{NL}}}\!\left(\! r,h \!\right)&= \exp \left( { - \int_0^{{r_2}} {2\pi {S_n}{\lambda _{\rm{s}}}{\Pr} _k^{\rm{L}}\left( u,h \right)u{\rm{d}}u} } \right)\nonumber\\
  &\times \exp \left( { - \int_0^r {2\pi {S_n}{\lambda _{\rm{s}}} \left({1 - {\Pr} _k^{\rm{L}}\left( u,h \right)} \right)u{\rm{d}}u} } \right) \nonumber\\
  &\times \left( {1 \!-\! {\Pr} _k^{\rm{L}}\left( r,h \right)} \right) 2\pi r{S_n}{\lambda _{\rm{s}}},\; {{d_{k \!-\! 1}(h)} \!<\! r \!\le\! {d_k}(h)},
  \end{align}
  where ${r_{1}} = \mathop {\arg }\limits_{{r_{1}}} \left\{ {{\zeta ^{{\rm{NL}}}}({r_1,h}) = \zeta _k^{\rm{L}}(r,h)} \right\}$ and ${r_2} = \mathop {\arg }\limits_{{r_2}} \left\{ {{\zeta ^{\rm{L}}}({r_2,h}) = \zeta _k^{{\rm{NL}}}(r,h)} \right\}$.

  $Proof:$ See Appendix A.\hfill $\blacksquare$


  To specify (\ref{equ:2}) and (\ref{equ:3}), we further have
  \begin{align}
  \label{equ:4}
  \Pr \!\left[\! {\frac{{P\zeta _k^{\rm{L}}\left(\! r,h \!\right)g}}{{{\sigma ^{\rm{2}}} + {I_Z}}}\!>\! \delta } \!\right]\! \!=\! \exp ( - {\textstyle{{\delta {\sigma ^2}} \over {P\zeta _k^{\rm{L}}\left( r,h \right)}}}){\mathscr{L} _{I_Z}}\!\left(\! {{\textstyle{\delta  \over {P\zeta _k^{\rm{L}}\left( r,h \right)}}}} \!\right),\\
  \label{equ:5}
  \Pr \!\left[\! {\frac{{P\zeta _k^{{\rm{NL}}}\!\left(\! r,h \!\right)\!g}}{{{\sigma ^{\rm{2}}} + {I_Z}}} \!>\! \delta } \!\right]\! \!= \! \exp (\! -\! {\textstyle{{\delta {\sigma ^2}} \over {P\zeta _k^{{\rm{NL}}}\left(\! r,h\! \right)}}}){\mathscr{L} _{I_Z}}\!\!\left(\! {{\textstyle{\delta  \over {P\zeta _k^{{\rm{NL}}}\left(\! r,h \!\right)}}}} \!\right)\!,
  \end{align}
  where $\mathscr{L}_{I_Z}(\cdot)$ is the Laplace transform of ${I_Z}$. We note that $I_Z$ at the typical UE that requests the $n$-th file comes from two independently portions: 1) $I_{Z1}$, caused by the SBSs in other tiers which locate in the entire area of the network, and 2) $I_{Z2}$, caused by the SBSs in the $n$-th tier whose distances with the typical UE are larger than $l$. Based on the observation, $I_Z = I_{Z1}+I_{Z2}$. Going forward, $Lemma$ 1 below computes ${\mathscr{L}}_{I_Z}\left( {{\textstyle{\delta  \over {P\zeta_k^{\rm{L}} \left( r,h \right)}}}} \right)$ in (\ref{equ:4}) and ${\mathscr{L}}_{I_Z}\left( {{\textstyle{\delta  \over {P\zeta_k^{\rm{NL}} \left( r,h \right)}}}} \right)$ in (\ref{equ:5}).

  $Lemma$ 1:
  \begin{align}
&{\mathscr{L} _{I_Z}}\left( {{\textstyle{\delta  \over {P\zeta_k \left( r,h \right)}}}} \right)\!=\! {E_{I_Z}}\left[ {\exp \left( { - {\textstyle{{\delta I_Z} \over {P{\zeta_k }\left( r,h \right)}}}} \right)} \right] \nonumber\\
&\!=\! {E_{{I_{Z1}}}}\left[ {\exp \left( { - {\textstyle{{\delta {I_{Z1}}} \over {P{\zeta_k}\left( r,h \right)}}}} \right)} \right] {E_{{I_{Z2}}}}\left[ {\exp \left( { - {\textstyle{{\delta {I_{Z2}}} \over {P{\zeta_k }\left( r,h \right)}}}} \right)} \right]\!.\!
\end{align}

  For LoS,
  \begin{align}
  \label{equ:11}
  &{E_{{I_{Z1}}}}\left[ {\exp \left( { - {\textstyle{{\delta {I_{Z1}}} \over {P{\zeta_k ^{\rm{L}}}\left( r,h \right)}}}} \right)} \right] = \exp \left\{ { - 2\pi \sum\limits_{i = 1,i \ne n}^N {\Pr \left( {{A_i}} \right){S_i}{\lambda _{\rm{s}}}}} \right. \nonumber \\
  &\times \left[ {\int_0^\infty  {{\textstyle{{{{\Pr }^{\rm{L}}}\left( {u },h \right)u} \over {1 + {\zeta_k ^{\rm{L}}}\left( r,h \right){{(\delta {\zeta ^{\rm{L}}}\left( u,h \right))}^{ - 1}}}}}} {\rm{d}}u} \right. \nonumber \\
  &\left. {\left. { + \int_0^\infty  {{\textstyle{{\left[ {1 - {{\Pr }^{\rm{L}}}\left( {u },h \right)} \right]u} \over {1 + {\zeta_k ^{\rm{L}}}\left( r,h \right){{(\delta {\zeta ^{\rm{NL}}}\left( u,h \right))}^{ - 1}}}}}{\rm{d}}u} } \right]} \right\},
\end{align}
  \begin{align}
  \label{equ:12}
   &{E_{{I_{Z2}}}}\left[ {\exp \left( { - {\textstyle{{\delta {I_{Z2}}} \over {P{\zeta_k ^{\rm{L}}}\left( r,h \right)}}}} \right)} \right]\nonumber\\
   &= \exp \left\{ { - 2\pi \Pr \left( {{A_n}} \right){S_n}{\lambda _{\rm{s}}}\left[ {\int_r^\infty  {{\textstyle{{{{\Pr }^{\rm{L}}}\left( {u },h \right)u} \over {1 + {\zeta_k ^{\rm{L}}}\left( r,h \right){{(\delta {\zeta ^{\rm{L}}}\left( u,h \right))}^{ - 1}}}}}} {\rm{d}}u} \right.} \right.\nonumber\\
&\left. {\left. { + \int_{{r_1}}^\infty  {{\textstyle{{\left[ {1 - {{\Pr }^{\rm{L}}}\left( {u },h \right)} \right]u} \over {1 + {\zeta_k ^{\rm{L}}}\left( r,h \right){{(\delta {\zeta ^{\rm{NL}}}\left( u,h \right))}^{ - 1}}}}}} {\rm{d}}u} \right]} \right\}.
\end{align}

  For NLoS,
  \begin{align}
  \label{equ:13}
  &{E_{{I_{Z1}}}}\left[ {\exp \left( { - {\textstyle{{\delta {I_{Z1}}} \over {P{\zeta_k ^{\rm{NL}}}\left( r,h \right)}}}} \right)} \right] = \exp \left\{ { - 2\pi \sum\limits_{i = 1,i \ne n}^N {\Pr \left( {{A_i}} \right){S_i}{\lambda _{\rm{s}}}}} \right. \nonumber \\
  &\times \left[ {\int_0^\infty  {{\textstyle{{{{\Pr }^{\rm{L}}}\left( {u },h \right)u} \over {1 + {\zeta_k ^{\rm{NL}}}\left( r,h \right){{(\delta {\zeta ^{\rm{L}}}\left( u,h \right))}^{ - 1}}}}}} {\rm{d}}u} \right. \nonumber \\
  &\left. {\left. { + \int_0^\infty  {{\textstyle{{\left[ {1 - {{\Pr }^{\rm{L}}}\left( {u },h \right)} \right]u} \over {1 + {\zeta_k ^{\rm{NL}}}\left( r,h \right){{(\delta {\zeta ^{\rm{NL}}}\left( u,h \right))}^{ - 1}}}}}{\rm{d}}u} } \right]} \right\},
\end{align}
  \begin{align}
  \label{equ:14}
   &{E_{{I_{Z2}}}}\left[ {\exp \left( { - {\textstyle{{\delta {I_{Z2}}} \over {P{\zeta_k ^{{\rm{NL}}}}\left( r,h \right)}}}} \right)} \right] \nonumber\\
   &= \exp \left\{ { - 2\pi \Pr \left( {{A_n}} \right){S_n}{\lambda _{\rm{s}}}\left[ {\int_{{r_2}}^\infty  {{\textstyle{{{{\Pr }^{\rm{L}}}\left( {u },h \right)u} \over {1 + {\zeta_k ^{{\rm{NL}}}}\left( r,h \right){{(\delta {\zeta ^{\rm{L}}}\left( u,h \right))}^{ - 1}}}}}} {\rm{d}}u} \right.} \right. \nonumber\\
  &\left. {\left. { + \int_r^\infty  {{\textstyle{{\left[ {1 - {{\Pr }^{\rm{L}}}\left( {u },h \right)} \right]u} \over {1 + {\zeta_k ^{{\rm{NL}}}}\left( r,h \right){{(\delta {\zeta ^{\rm{NL}}}\left( u,h \right))}^{ - 1}}}}}{\rm{d}}u} } \right]} \right\}.
\end{align}

  $Proof:$ See Appendix B. \hfill $\blacksquare$

  Let $\Pr \left( {A_n} \right){S_n}{\lambda _{\rm{s}}}$ be the density of active SBSs in the $n$-th tier. Eqn. (\ref{equ:6}) implies that the density of the active SBSs increases as the UE density goes up. From $Theorem$ 1 and $Lemma$ 1, the increase of the UE density degrades its SDP.


  Finally, considering the request probabilities of all $N$ files, we obtain the average SDP that the UEs can successfully download all possible files as
  \begin{equation}
  \overline {\Pr} = \sum\limits_{n = 1}^N {{Q_n} \Pr \left( {{D_n}} \right)} .
  \end{equation}

\subsection{A 3GPP Path Loss Model for TUs}
  For the TUs, we show a path loss function adopted by 3GPP \cite{TR_36828}, i.e.,
  \begin{equation}
  \label{equ:7}
  {\zeta _{\rm{t}}}\left( r,h_1 \right) = \left\{ {\begin{array}{*{20}{l}}
{A_{\rm{t}}^{\rm{L}}{l^{{\rm{ - \alpha }}_{\rm{t}}^{\rm{L}}}},\;\;\;\;\;{\text{with}}\;\Pr _{\rm{t}}^{\rm{L}}\left( r,h_1 \right)}\\
{A_{\rm{t}}^{{\rm{NL}}}{l^{{\rm{ - \alpha }}_{\rm{t}}^{{\rm{NL}}}}},\;{\text{with}}\;\left( {1 - \Pr _{\rm{t}}^{\rm{L}}\left( r,h_1 \right)} \right)}
\end{array}} \right.,
  \end{equation}
   together with a linear LoS probability function also adopted by 3GPP, i.e.,
  \begin{equation}
  \label{equ:8}
  {{\rm{Pr}}_{\rm{t}}^{\rm{L} }}\left( r,h_1 \right) = \left\{ {\begin{array}{*{20}{c}}
{1 - {\textstyle{l \over {{l_0}}}},}&\;\;\;\;\;{0 < r \le {d_{\rm{T}}}}\\
{0,}&{r > {d_{\rm{T}}}}
\end{array}} \right.,
  \end{equation}
  where $l_0$ is the cut-off distance of the LoS link, and ${d_{\rm{T}}}\triangleq d_1(h_1)= \sqrt {{l_0^2} - {h_1^2}}$.

  We remark that the path loss model in (\ref{equ:7}) and (\ref{equ:8}) is a special case of the general path loss model in (\ref{equ:1}) with the following substitutions: $N = 2$, $\zeta _{{\rm{t}},1}^{\rm{L}}\left( r,h_1 \right) = \zeta _{{\rm{t}},2}^{\rm{L}}\left( r,h_1 \right) = A_{\rm{t}}^{\rm{L}}{l^{{\rm{ - \alpha }}_{\rm{t}}^{\rm{L}}}}$, $\zeta _{{\rm{t}},1}^{{\rm{NL}}}\left( r,h_1 \right) = \zeta _{{\rm{t}},2}^{{\rm{NL}}}\left( r,h_1 \right) = A_{\rm{t}}^{{\rm{NL}}}{l^{{\rm{ - \alpha }}_{\rm{t}}^{{\rm{NL}}}}}$, $\Pr _{\rm{t},1}^{\rm{L}}\left( r,h_1 \right) = 1 - {\textstyle{l \over {{l_0}}}}$ and $\Pr _{{\rm{t}},2}^{\rm{L}}\left( r,h_1 \right) = 0$.

  For the 3GPP path loss model, by $Theorem$ 1, the probability that the typical TU successfully receives the requested $n$-th file from the associated $n$-th tier SBS becomes
  \begin{align}
{{\Pr}_{\rm{t}}}\left( {{D_n}} \right) &\!=\! \sum\limits_{k = 1}^2 {\left( {T_{{\rm{t}},k}^{\rm{L}} + T_{{\rm{t}},k}^{{\rm{NL}}}} \right)} \nonumber \\
  &\!=\! \int_0^{{d_{\rm{T}}}} {\exp ( - {\textstyle{{\delta {\sigma ^2}{l^{\alpha _{\rm{t}}^{\rm{L}}}}} \over {PA_{\rm{t}}^{\rm{L}}}}}){\mathscr{L} _{I_Z}}\left( {{\textstyle{{\delta {l^{\alpha _{\rm{t}}^{\rm{L}}}}} \over {PA_{\rm{t}}^{\rm{L}}}}}} \right)} f_{{\rm{t}},1}^{\rm{L}}\left( r,h_1 \right){\rm{d}}r \!+\!0 \nonumber \\
  &\!+\! \int_0^{{d_{\rm{T}}}} {\exp ( - {\textstyle{{\delta {\sigma ^2}{l^{\alpha _{\rm{t}}^{{\rm{NL}}}}}} \over {PA_{\rm{t}}^{{\rm{NL}}}}}}){\mathscr{L} _{I_Z}}\left( {{\textstyle{{\delta {l^{\alpha _{\rm{t}}^{{\rm{NL}}}}}} \over {PA_{\rm{t}}^{{\rm{NL}}}}}}} \right)} f_{{\rm{t}},1}^{{\rm{NL}}}\left( r,h_1 \right){\rm{d}}r \nonumber \\
 &\!+\! \int_{{d_{\rm{T}}}}^\infty  {\exp ( - {\textstyle{{\delta {\sigma ^2}{l^{\alpha _{\rm{t}}^{{\rm{NL}}}}}} \over {PA_{\rm{t}}^{{\rm{NL}}}}}}){\mathscr{L} _{I_Z}}\left( {{\textstyle{{\delta {l^{\alpha _{\rm{t}}^{{\rm{NL}}}}}} \over {PA_{\rm{t}}^{{\rm{NL}}}}}}} \right)} f_{{\rm{t}},2}^{{\rm{NL}}}\left( r,h_1 \right){\rm{d}}r \nonumber \\
 &\!=\! T_{{\rm{t}},1}^{\rm{L}} + T_{{\rm{t}},1}^{{\rm{NL}}} + T_{{\rm{t}},2}^{{\rm{NL}}},
 \end{align}
  where   $T_{{\rm{t}},2}^{{\rm{L}}}=0$, because $\Pr _{\rm{t}}^{\rm{L}}\left( r,h_1 \right)=0$ when $r > {d_{\rm{T}}}$. For LoS, $f_{{\rm{t}},1}^{\rm{L}}\left( r,h_1 \right)$ and ${\mathscr{L} _{I_Z}}\left( {{\textstyle{{\delta {l^{\alpha _{\rm{t}}^{\rm{L}}}}} \over {PA_{\rm{t}}^{\rm{L}}}}}} \right)$ are calculated by (\ref{equ:9}), (\ref{equ:11}) and (\ref{equ:12}).
  For NLoS, $f_{{\rm{t}},1}^{\rm{NL}}\left( r,h_1 \right)$, $f_{{\rm{t}},2}^{\rm{NL}}\left( r,h_1 \right)$ and ${\mathscr{L} _{I_Z}}\left( {{\textstyle{{\delta {l^{\alpha _{\rm{t}}^{\rm{NL}}}}} \over {PA_{\rm{t}}^{\rm{NL}}}}}} \right) $ are calculated by (\ref{equ:10}), (\ref{equ:13}) and (\ref{equ:14}).

\subsection{A 3GPP Path Loss Model for UAVs}

  For the UAVs, we consider a path loss function adopted by 3GPP \cite{TR_36777}, i.e.,
  \begin{equation}
  \label{equ:15}
  {\zeta _{\rm{a}}}\left( r,h_2 \right) = \left\{ {\begin{array}{*{20}{l}}
{A_{\rm{a}}^{\rm{L}}{l^{{\rm{ - \alpha }}_{\rm{a}}^{\rm{L}}}},\;\;\;\;\;{\text{with}}\;\Pr _{\rm{a}}^{\rm{L}}\left( r,h_2 \right)}\\
{A_{\rm{a}}^{{\rm{NL}}}{l^{{\rm{ - \alpha }}_{\rm{a}}^{{\rm{NL}}}}},\;{\text{with}}\;\left( {1 - \Pr _{\rm{a}}^{\rm{L}}\left( r,h_2 \right)} \right)}
\end{array}} \right.,
  \end{equation}
  together with a LoS probability function also adopted by 3GPP \cite{TR_36777}, i.e.,
  \begin{equation}
  \label{equ:16}
  {{\Pr}_{\rm{a}}^{\rm{L}}}\left( r,h_2 \right) \!=\! \!\left\{\! {\begin{array}{*{20}{c}}
{1,}&{0 \!<\! r \!\le\! {d_{\rm{A}}}}\\
{{\textstyle{{{d_{\rm{A}}}} \over r}} \!+\! \exp\!\left(\! {{\textstyle{{ - r} \over {{p_1}}}}} \!\right)\!\left( {1\! -\! {\textstyle{{{d_{\rm{A}}}} \over r}}} \right)\!,\!}&\hspace{-0.5cm}{r \!>\! {d_{\rm{A}}}}
\end{array}} \right.\!,\!
  \end{equation}
  where
  \begin{equation}
  {p_1} = 233.98{\log _{10}}\left( {h_{\rm{AU}}} \right) - 0.95,
  \end{equation}
  \begin{equation}
  {d_{\rm{A}}} \!\triangleq\! d_1(h_2)\!=\! \max \left( {294.05{{\log }_{10}}\left( {h_{\rm{AU}}} \right) \!-\! 432.94,{\kern 1pt} 18} \right).
  \end{equation}

  From this 3GPP channel model, the applicability range in terms of UAV height is $22.5 {\rm{m}}\leq{h_{\rm{AU}}}\leq300 {\rm{m}}$~\cite{TR_36777}.
  Note that the path loss model in (\ref{equ:15}) and (\ref{equ:16}) is also a special case of (\ref{equ:1}) with the following substitutions: $N = 2$, $\zeta _{{\rm{a}},1}^{\rm{L}}\left( r,h_2 \right) = \zeta _{{\rm{a}},2}^{\rm{L}}\left( r,h_2 \right) = A_{\rm{a}}^{\rm{L}}{l^{{\rm{ - \alpha }}_{\rm{a}}^{\rm{L}}}}$, $\zeta _{{\rm{a}},1}^{{\rm{NL}}}\left( r,h_2 \right) = \zeta _{{\rm{a}},2}^{{\rm{NL}}}\left( r,h_2 \right) = A_{\rm{a}}^{{\rm{NL}}}{l^{{\rm{ - \alpha }}_{\rm{a}}^{{\rm{NL}}}}}$, $\Pr _{{\rm{a}},1}^{\rm{L}}\left( r,h_2 \right) = 1$ and $\Pr _{{\rm{a}},2}^{\rm{L}}\left( r,h_2 \right) = {\textstyle{{{d_{\rm{A}}}} \over r}} + \mathrm{exp}\left( {{\textstyle{{ - r} \over {{p_1}}}}} \right)\left( {1 - {\textstyle{{{d_{\rm{A}}}} \over r}}} \right)$. 

  From $Theorem$ 1, the probability that the typical UAV successfully receives the requested $n$-th file can be given by
  \begin{align}
{{\Pr} _{\rm{a}}}\left( {{D_n}} \right) &\!= \! \int_0^{{d_{\rm{A}}}} {\exp ( - {\textstyle{{\delta {\sigma ^2}{l^{\alpha _{\rm{a}}^{\rm{L}}}}} \over {PA_{\rm{a}}^{\rm{L}}}}}){\mathscr{L} _{I_Z}}\left( {{\textstyle{{\delta {l^{\alpha _{\rm{a}}^{\rm{L}}}}} \over {PA_{\rm{a}}^{\rm{L}}}}}} \right)} f_{{\rm{a}},1}^{\rm{L}}\left( r,h_2 \right){\rm{d}}r \nonumber \\
 &\!+\! \int_{{d_{\rm{A}}}}^\infty  {\exp ( - {\textstyle{{\delta {\sigma ^2}{l^{\alpha _{\rm{a}}^{\rm{L}}}}} \over {PA_{\rm{a}}^{\rm{L}}}}}){\mathscr{L} _{I_Z}}\left( {{\textstyle{{\delta {l^{\alpha _{\rm{a}}^{\rm{L}}}}} \over {PA_{\rm{a}}^{\rm{L}}}}}} \right)} f_{{\rm{a}},2}^{\rm{L}}\left( r,h_2 \right){\rm{d}}r\!+\!0 \nonumber \\
 &\!+\! \int_{{d_{\rm{A}}}}^\infty  {\exp ( - {\textstyle{{\delta {\sigma ^2}{l^{\alpha _{\rm{a}}^{{\rm{NL}}}}}} \over {PA_{\rm{a}}^{{\rm{NL}}}}}}){\mathscr{L} _{I_Z}}\left( {{\textstyle{{\delta {l^{\alpha _{\rm{a}}^{{\rm{NL}}}}}} \over {PA_{\rm{a}}^{{\rm{NL}}}}}}} \right)} f_{{\rm{a}},2}^{{\rm{NL}}}\left( r,h_2 \right){\rm{d}}r \nonumber \\
 &\!=\! T_{{\rm{a}},1}^{\rm{L}} + T_{{\rm{a}},2}^{\rm{L}} + T_{{\rm{a}},2}^{{\rm{NL}}},
\end{align}
  where  $T_{{\rm{a}},1}^{{\rm{NL}}}=0$, because $\Pr _{\rm{a}}^{\rm{NL}}\left( r,h_2 \right)=0$ when $0 < r \le {d_{\rm{A}}}$. For LoS, $f_{{\rm{a}},1}^{\rm{L}}\left( r,h_2 \right)$, $f_{{\rm{a}},2}^{\rm{L}}\left( r,h_2 \right)$ and ${\mathscr{L} _{I_Z}}\left( {{\textstyle{{\delta {l^{\alpha _{\rm{a}}^{\rm{L}}}}} \over {PA_{\rm{a}}^{\rm{L}}}}}} \right)$ are calculated by (\ref{equ:9}), (\ref{equ:11}) and (\ref{equ:12}).
  For NLoS, $f_{{\rm{a}},2}^{\rm{NL}}\left( r,h_2 \right)$ and ${\mathscr{L} _{I_Z}}\left( {{\textstyle{{\delta {l^{\alpha _{\rm{a}}^{\rm{NL}}}}} \over {PA_{\rm{a}}^{\rm{NL}}}}}} \right) $ are calculated by (\ref{equ:10}), (\ref{equ:13}) and (\ref{equ:14}).

\section{Optimized Caching Probabilities }

  In dynamic on-off architecture, (\ref{equ:6}) shows that Pr($A_n$) is a function of the ratio ${{{Q_n}{\lambda _{\rm{u}}}}/{{S_n}{\lambda _{\rm{s}}}}}$. Since the SBS density is much higher than the UE density in this architecture, i.e., ${\lambda _{\rm{s}}} >> {\lambda _{\rm{u}}}$, Pr($A_n$) can be approximated as \cite{Probabilistic_SC_Caching}
  \begin{equation}
  \label{equ:25}
  \Pr \left( {{A_n}} \right) \approx {\textstyle{\frac{{{Q_n}{\lambda _{\rm{u}}}}}{{{S_n}{\lambda _{\rm{s}}}}}}}.
  \end{equation}

  Since both TU and UAV can be regarded as the typical user, the average SDP is given by
  \begin{equation}
  {\overline{\Pr}}  = \sum\limits_{n = 1}^N Q_n {\left( {\textstyle{\frac{{{\lambda _{{\rm{TU}}}}}}{{{\lambda _{\rm{u}}}}}}{{\Pr }_{\rm{t}}}\left( {{D_n}} \right) + \frac{{{\lambda _{\rm{AU}}}}}{{{\lambda _{\rm{u}}}}}{{\Pr }_{\rm{a}}}\left( {{D_n}} \right)} \right)} .
  \end{equation}

  Consider that the SBS density approaches infinity. In this scenario, the downlink transmission from the typical SBS to the typical UE is dominantly characterized by the LoS path loss. In this context, we derive the asymptotic performance of $\overline{\Pr}$ as follows.

  $Theorem$ 2: In a noise-free case where the SBS density goes to infinity, i.e., $\lambda_{\rm{s}}  \to  + \infty $  and ${\sigma ^2} = 0$, the average SDP is given by
  \begin{small}
  \begin{align}
  \label{equ:23}
  \overline {\Pr }  &= \sum\limits_{n = 1}^N {{Q_n}{S_n}\left[ {\int_0^{{d_{\rm{T}}}} {{G_n}\exp \left( { - \pi {S_n}{\lambda _{\rm{s}}}{r^2}} \right)} {\rm{d}}r} \right.} \nonumber \\
  &\left. { + \int_0^{{d_{\rm{A}}}} {{H_n}\exp \left( { - \pi {S_n}{\lambda _{\rm{s}}}{r^2}} \right)} {\rm{d}}r} \right],
  \end{align}
  \end{small}
where
  \begin{small}
  \begin{align}
  \label{equ:17}
  {G_n}\!=\! {{{{\lambda _{\rm{TU}}}2\pi {\lambda _{\rm{s}}}r} \over {{\lambda _{\rm{u}}}}}} \exp \!\left(\! {\sum\limits_{i \!=\! 1,i \!\ne\! n}^N \!{{Q_i}} B\!\left(\! {r,h_1} \!\right)\! +\! {Q_n}C\left( {r,h_1} \!\right)\!} \right), \\
  \label{equ:18}
  {H_n}\! =\! {{{{\lambda _{\rm{AU}}}2\pi {\lambda _{\rm{s}}}r} \over {{\lambda _{\rm{u}}}}}} \exp \!\left(\! {\sum\limits_{i \!=\! 1,i \!\ne\! n}^N \!{{Q_i}} E\!\left(\! {r,h_2} \!\right) \!+\! {Q_n}F\left( {r,h_2} \!\right)\!} \right)\!.\!
  \end{align}
  \end{small}

  In (\ref{equ:17}) and (\ref{equ:18}), $B\left( {r,h_1} \right)$, $C\left( {r,h_1} \right)$, $E\left( {r,h_2} \right)$ and $F\left( {r,h_2} \right)$ are given by (\ref{equ:b}), (\ref{equ:c}), (\ref{equ:e}) and (\ref{equ:f}) respectively in Appendix C.

  $Proof$: See Appendix C. \hfill $\blacksquare$

  From $Theorem$ 1, $Lemma$ 1 and $Theorem$ 2, we show that the SDPs of TUs and UAVs are correlated
  , i.e., the increase of the TU (or UAV) density degrades the SDP performance of all UEs, including TUs and UAVs.
  As a result, we cannot separately maximize the average SDP for each type of UEs. Instead, we jointly maximize the average SDP of TUs and UAVs.


  According to  $Theorem$ 2, we can formulate the optimization problem of maximizing $\overline {\Pr}$ as
  \begin{align}
{\text{P1}}: &\mathop {\rm{max}}\limits_{{S_n}\left| {_{n = 1}^N} \right.} \;\overline {\Pr} \nonumber\\
&= {\mathop {\rm{max}}\limits_{{S_n}\left| {_{n = 1}^N} \right.} \;\sum\limits_{n = 1}^N {{Q_n}{S_n}\left[ {\int_0^{{d_{\rm{T}}}} {{G_n}\exp \left( { - \pi {S_n}{\lambda _{\rm{s}}}{r^2}} \right)} {\rm{d}}r} \right.}}\nonumber \\
&{\left. { + \int_0^{{d_{\rm{A}}}} {{H_n}\exp \left( { - \pi {S_n}{\lambda _{\rm{s}}}{r^2}} \right)} {\rm{d}}r} \right]} \nonumber \\
&{\rm{s}}{\rm{.t}}{\rm{.}}\;\;\sum\limits_{n = 1}^N {{S_n} \leq } M \nonumber\\
&\;\;\;\;\;\;\;0 \le {S_n} \le 1,\;n = 1, \cdots ,N.
\end{align}

  Note that P1 is a non-convex optimization problem. To cope with this problem, we transform P1 into $N$ sub-problems and solve it in parallel.
  Adopting the partial Lagrangian, we have
  \begin{align}
  &L(S_1, S_2, \cdots, S_N, \gamma)\nonumber \\
  &= \sum\limits_{n = 1}^N {{Q_n}{S_n}\left[ {\int_0^{{d_{\rm{T}}}} {{{G}_n}\exp \left( { - \pi {S_n}{\lambda _s}{r^2}} \right)} {\rm{d}}r} \right.} \nonumber \\
  &\left. { \!+\! \int_0^{{d_{\rm{A}}}} {{{H}_n}\exp \left( {\! -\! \pi {S_n}{\lambda _s}{r^2}} \right)} {\rm{d}}r} \right] + \gamma \left( {\sum\limits_{n = 1}^N {{S_n} \!-\! } M} \!\right)\!,\!
\end{align}
  where $\gamma M$ is constant and ${0 \!\le\! {S_n} \!\le\! 1,\;n = 1, \cdots ,N}$. In the following, we opt to optimize each individual sub-problem.

  First, the partial Lagrangian of the $n$-th sub-problem with respect to $S_n$ is given by
  \begin{align}
  \label{equ:19}
L\left( {{S_n},\gamma } \right) &\!=\! \int_0^{{d_{\rm{T}}}} {{Q_n}{S_n}{G_n}\exp \left( { - \pi {S_n}{\lambda _{\rm{s}}}{r^2}} \right)} {\rm{d}}r \nonumber \\
  &\!+\! \int_0^{{d_{\rm{A}}}} {{Q_n}{S_n}{H_n}\exp \left( { - \pi {S_n}{\lambda _{\rm{s}}}{r^2}} \right)} {\rm{d}}r \!+\! \gamma {S_n}.
  \end{align}

  Second, by some mathematical manipulation, we have
  \begin{align}
  \label{equ:20}
&\textstyle{\frac{{\partial L \left( {{S_n},\gamma } \right)}}{{\partial {S_n}}}} \\
 &\!=\!\int_0^{{d_{\rm{T}}}} \textstyle{{{Q_n}{G_n}\exp \left( { \!-\! W{S_n}} \right) \!-\! } W{Q_n}{S_n}{G_n}\exp \left( { \!-\! W{S_n}} \right)}{\rm{d}}r \nonumber \\ \nonumber
 &\!+\!\int_0^{{d_{\rm{A}}}} \textstyle{\!{{Q_n}{H_n}\exp \left( { \!-\! W{S_n}} \right)}  \!-\! W{Q_n}{S_n}{H_n}\exp \left( { \!-\! W{S_n}} \right)}{\rm{d}}r \!+\! \gamma \!,\!
\end{align}
  where $W =  \pi {\lambda _{\rm{s}}}{r^2}$.

  Here, we discuss the key steps of optimizing the caching probabilities in Algorithm 1. With $\frac{{\partial L \left( {{S_n},\gamma } \right)}}{{\partial {S_n}}}=0$, we can get the extreme point $S_n$. Let ${{\bf{S}}_n} = \left[ {S_n},0,1\right] $ if $0 \le {S_n} \le 1$. Otherwise, let ${{\bf{S}}_n} = \left[0, 1\right] $. Then, we identify the element in $\mathbf{S}_n$ that maximizes the $L\left( {{S_n},\gamma } \right) $ in (\ref{equ:19}) as the optimized caching probability ${S_n}$. After $N$ iterations, we get $\sum\limits_{i = 1}^N {{{S }_i}} $.
  If $\sum\limits_{i = 1}^N {{{S }_i}} = M$, we obtain the optimized caching probability $\left\{ {{S_1}, {S_2}, \cdots, {S_N}} \right\}$. Otherwise, the dual variable $\gamma$ in (\ref{equ:20}) is updated by ${\gamma _{i + 1}} =  {{\gamma _i} + \left( {{S_1} + {S_2} +  \cdots  + {S_N} - M} \right)\varphi }$, where $\varphi$ is the step size.
\begin{algorithm}
\caption{Optimized Caching Probabilities in the Dynamic On-Off Architecture}
\begin{algorithmic}[1]
    \STATE Initial $\gamma$, $\varphi$, $M$ and $N$
    \REPEAT
        \FOR{$n=1:1:N$}
        \STATE Compute $S_n$ from $\frac{{\partial L\left( {{S_n},\gamma } \right)}}{{\partial {S_n}}} = 0$ in (\ref{equ:20})
        \IF{$0 \le {S_n} \le 1$}
            \STATE Set ${{\bf{S}}_n} =\left[{S_n}, 0, 1\right]$
        \ELSE
            \STATE Set ${{\bf{S}}_n} =\left[0, 1\right]$
        \ENDIF
        \STATE Plug $S'_n \in {{\bf{S}}_n}$ into (\ref{equ:19})
        \STATE Choose ${S_n} = \arg \mathop {\max }\limits_{S'_n} L\left( {{S'_n},\gamma } \right)$
        \ENDFOR
        \STATE Get $S=\sum\limits_{i = 1}^N {{S_i}} $, then update $\gamma$
  \UNTIL{$S = M$}
\end{algorithmic}
\end{algorithm}

  The analytical results in Section V build upon an assumption that both UAVs and TUs have the same content request probability. The following remark briefly discusses the case where the two tiers of UEs have different request probabilities.

  $Remark$ 1: Let $Q_n^{\rm{T}}$ and $Q_n^{\rm{A}}$ be the request probabilities of the $n$-th file for TUs and UAVs respectively. From (34), the weighted sum request probability becomes ${Q_n} = {\textstyle{{{\lambda _{{\rm{TU}}}}} \over {{\lambda _{\rm{u}}}}}}Q_n^{\rm{T}} + {\textstyle{{{\lambda _{{\rm{AU}}}}} \over {{\lambda _{\rm{u}}}}}}Q_n^{\rm{A}}$, and $\overline {\Pr }  = \sum\limits_{n = 1}^N {\left( {\frac{{Q_n^{\rm{T}}{\lambda _{{\rm{TU}}}}}}{{{\lambda _{\rm{u}}}}}{{\Pr }_{\rm{t}}}\left( {{D_n}} \right) + \frac{{Q_n^{\rm{A}}{\lambda _{{\rm{AU}}}}}}{{{\lambda _{\rm{u}}}}}{{\Pr }_{\rm{a}}}\left( {{D_n}} \right)} \right)}$. It can be seen that the optimization of SDP in this case largely follows (36)-(42) in this Section. In addition, the SBSs are more likely to cache the file requested by the UEs with better channel quality, when the TUs and UAVs have the same request probability for different files.

\vspace{-0.1in}
  \section{Analysis on Network Parameters under UCS and PCS}
  In this section, we analyze the performance limits of the average SDP for the UCS and PCS under a single-slope path loss model \cite{Probabilistic_SC_Caching}, respectively, where the path loss of the channel  from an SBS to an UE is modeled as $l^{-\alpha}$ with $\alpha$ denoting the path loss exponent. In a noise-free case, the average SDP is given by (\ref{equ:21}) (see the top of next page).
  \begin{figure*}[ht]
  \begin{align}
  \label{equ:21}
\overline{\Pr}= &\sum\limits_{n = 1}^N {{Q_n}\int_0^\infty  {{\mathscr{L} _{I_Z}}\left( {{\textstyle{{\delta {l^\alpha }} \over P}}} \right)} f\left( r \right){\rm{d}}r} = \sum\limits_{n = 1}^N {{Q_n}\int_0^\infty  {\exp \left( { - 2\pi \sum\limits_{i = 1,i \ne n}^N {\Pr \left( {{A_i}} \right){S_i}{\lambda _{\rm{s}}}} \int_0^\infty  {{\textstyle{u \over {1 + {l^{-\alpha} }{\delta ^{ - 1}}{{\left( {\sqrt {{u^2} + {h^2}} } \right)}^\alpha }}}}} {\rm{d}}u} \right)} } \nonumber \\
&\times \exp \left( { - 2\pi \Pr \left( {{A_n}} \right){S_n}{\lambda _{\rm{s}}}\int_r^\infty  {{\textstyle{{{u}} \over {1 + {l^{-\alpha} }{\delta ^{ - 1}}{{\left( {\sqrt {{u^2} + {h^2}} } \right)}^\alpha }}}}} {\rm{d}}u} \right) 2\pi {S_n}{\lambda _{\rm{s}}}r\exp \left( { - \pi {S_n}{\lambda _{\rm{s}}}{r^2}} \right){\rm{d}}r.
\end{align}
\hrulefill
\end{figure*}

  $Theorem$ 3: Consider a single-slope path loss model. When the SBS density $\lambda_{\rm{s}}$ is large enough, the average SDP is given by
  \begin{equation}
  \begin{array}{l}
\overline{\Pr} = \sum\limits_{n = 1}^N {\frac{{{Q_n}{S_n}\exp \left( { - \pi {h^2}{\lambda _{\rm{u}}}\mathcal{F}(\delta ,\alpha )} \right)}}{{{S_n} + \frac{{{\lambda _{\rm{u}}}}}{{{\lambda _{\rm{s}}}}}\mathcal{F}(\delta ,\alpha )}}}.
\end{array}
  \end{equation}

  For the PCS,
$\overline{\Pr} = \sum\limits_{n = 1}^M {\frac{{{Q_n}\exp \left( { - \pi {h^2}{\lambda _{\rm{u}}}\mathcal{F}(\delta ,\alpha )} \right)}}{{1 + \frac{{{\lambda _{\rm{u}}}}}{{{\lambda _{\rm{s}}}}}\mathcal{F}(\delta ,\alpha )}}}$.

  For the UCS,
$\overline{\Pr} = \frac{{\exp \left( { - \pi {h^2}{\lambda _{\rm{u}}}\mathcal{F}(\delta ,\alpha )} \right)}}{{1 + {\textstyle{{N{\lambda _{\rm{u}}}} \over {M{\lambda _{\rm{s}}}}}}\mathcal{F}(\delta ,\alpha )}}$.

  $Proof$: See Appendix D. \hfill $\blacksquare$

  From $Theorem$ 3, $Corollary$ 1 and $Corollary$ 2 below show the performance limits of the average SDPs under the PCS and UCS respectively.

  $Corollary$ 1: For $\lambda _{\rm{s}}  \to  + \infty $, the performance limits of the average SDPs under the PCS and UCS are given by respectively,
  \begin{align}
  \overline{\Pr} = \sum\limits_{n = 1}^M {{Q_n}\exp \left( { - \pi {h^2}{\lambda _{\rm{u}}}\mathcal{F}(\delta ,\alpha )} \right)},\; {\text{for the PCS}},\\
  \overline{\Pr} = \exp \left( { - \pi {h^2}{\lambda _{\rm{u}}}\mathcal{F}(\delta ,\alpha )} \right),\; {\text{for the UCS}}.
  \end{align}

  $Corollary$ 2: As the exponent $\beta  \to  + \infty $, the performances limits of the average SDPs for the PCS and UCS are given by respectively,
  \begin{align}
  \overline {\Pr} = \textstyle\frac{{\exp \left( { - \pi {h^2}{\lambda _{\rm{u}}}\mathcal{F}(\delta ,\alpha )} \right)}}{{1 + {\textstyle{{{\lambda _{\rm{u}}}} \over {{\lambda _{\rm{s}}}}}}\mathcal{F}(\delta ,\alpha )}},\; {\text{for the PCS}}\\
  \overline {\Pr}  = \textstyle\frac{{\exp \left( { - \pi {h^2}{\lambda _{\rm{u}}}\mathcal{F}(\delta ,\alpha )} \right)}}{{1 + {\textstyle{{N{\lambda _{\rm{u}}}} \over {M{\lambda _{\rm{s}}}}}}\mathcal{F}(\delta ,\alpha )}},\; {\text{for the UCS}}.
  \end{align}

  Based on $Theorem$ 3, $Corollary$ 1 and $Corollary$ 2, we have the following remarks.

  $Remark$ 2: From $Theorem$ 3, the average SDPs of both PCS and UCS increase as the SBS density $\lambda_{\rm{s}}$ increases. When $\lambda _{\rm{s}}  \to  + \infty $, $Corollary$ 1 shows that the UCS achieves a better average SDP than the PCS. In particular, the SDP of the PCS is affected by $\beta$ and $M$, while it is not the case for the UCS.

  $Remark$ 3: From $Theorem$ 3, as the cache size $M$ increases, the average SDPs of both PCS and UCS increase, and the performance gap for the two strategies narrows down. In particular, both strategies achieve the same SDP when $M = N$.

  $Remark$ 4: From $Theorem$ 3, the average SDPs of both PCS and UCS increase as $\beta$ gradually
  increases. In particular, the SDP of the PCS grows more rapidly. Based on $Corollary$ 2, the PCS outperforms the UCS when $\beta  \to  + \infty $.

  $Remark$ 5: From $Theorem$ 3, the average SDPs of both PCS and UCS become smaller as the height difference $h$ increases.

    \begin{table}[b]
  \centering
  \caption{The network parameters for TUs and UAVs}
  \begin{tabular}{|c|c|c|c|}
    \hline
    \multicolumn {2}{|c|}{TUs} & \multicolumn {2}{c|}{UAVs}  \\  \hline
    $A_{\rm{t}}^{\rm{L}}$       & $10^{ - 4.11}$  & $A_{\rm{a}}^{\rm{L}}$  & $10^{ - 3.692}$                                   \\  \hline
    $A_{\rm{t}}^{\rm{NL}}$      & $10^{ - 3.29}$  & $A_{\rm{a}}^{\rm{NL}}$ & $10^{ - 3.842}$                                   \\  \hline
    $\alpha_{\rm{t}}^{\rm{L}}$  & 2.09    & $\alpha_{\rm{a}}^{\rm{L}}$     & $2.225 - 0.05{\log_{10}}\left( {{h_{\rm{AU}}}} \right)$ \\  \hline
    $\alpha_{\rm{t}}^{\rm{NL}}$ & 3.75    & $\alpha_{\rm{a}}^{\rm{NL}}$    & $4.32 - 0.76{\log_{10}}\left( {{h_{\rm{AU}}}} \right)$  \\  \hline
    $\lambda_{\rm{TU}}$ & 150 ${\rm{TUs/km}^2}$ & $\lambda_{\rm{AU}}$  & 150 ${\rm{AUs/km}^2}$                             \\  \hline
    $h_{\rm{TU}}$             & 1.5m            & $h_{\rm{AU}}$        & 30m                                               \\  \hline
  \end{tabular}
  \end{table}

  \section{Numerical and Simulation Results}
  In this section, we use both numerical results and Monte Carlo simulation results to validate our analytical results. In the simulations, the performance is averaged over $10^5$ network deployments, where in each deployment SBSs and UEs are randomly distributed according to HPPPs with different densities. According to the 3GPP recommendations \cite{TR_36777}, \cite{LoS_NLoS} and \cite{TR_36828}, we use $h_{\rm{BS}} = 10{\rm{m}}$, $P = 24{\rm{dBm}}$, $\delta = -6{\rm{dB}}$. Other parameters for TUs and UAVs are listed in Table II.
  According to the applicability range of UAV height in \cite{TR_36777}, the UAV height is set to 30m.
  More specifically, we first focus on the average SDPs of different caching strategies in the UD SCNs. Second, we further investigate the impacts of the key network parameters, i.e., the SBS density, the cache size of cache memory, the exponent of Zipf distribution and the height of UAVs on the average SDP.
\vspace{-4mm}
\subsection{Impact of SBS Density}
   Fig. 2 compares the average SDPs $\overline {\Pr}$ versus the SBS density $\lambda_{\rm{s}}$ among the OCS, PCS and UCS for TUs and UAVs respectively. First, it can be seen that the numerical results match well with the simulation results in all scenarios. In the following, we focus on the analytical results only. Second, Fig. 2 shows that the OCS always outperforms the other two caching strategies.
   Third, in contrary to     \cite[Fig. 3]{Small_Cell_Caching} that uses the always-on architecture, we use a dynamic on-off architecture where an SBS is only active when it is required to serve the UEs. We show that the SDP increases with the increase of SBS density, while \cite{Small_Cell_Caching} showed that the SDP first increases and then drops down with the increase of SBS density.
   Fourth, we observe that the UCS can achieve a good performance as long as $\lambda_{\rm{s}}$ is large enough, in spite of a small caching probability. In this sense, it advocates caching some other files to further improve the saturated performance. These observations in Fig. 2 are in line with $Remark$ 2. The reasons are as follows:
  \begin{enumerate}
    \item when the SBS density is small, it is advisable to use the PCS because it smartly uses the limited number of SBSs to cache more popular files;
    \item when the SBS density is large, the coverage probability becomes saturated \cite{limit}. To be specific, the coverage probabilities are the same when $\lambda_{\rm{s}} = 10^5$ ${\rm{SBSs/km}^2}$ or $\lambda_{\rm{s}} = 10^6 \;{\rm{SBSs/km}^2}$. Hence, it is better to place the files randomly by the UCS than discarding less popular files by the PCS.
  \end{enumerate}
\begin{figure}[t]
\centering
\subfigure[$\overline \Pr$ of TUs]{
\includegraphics[height=0.69\columnwidth]{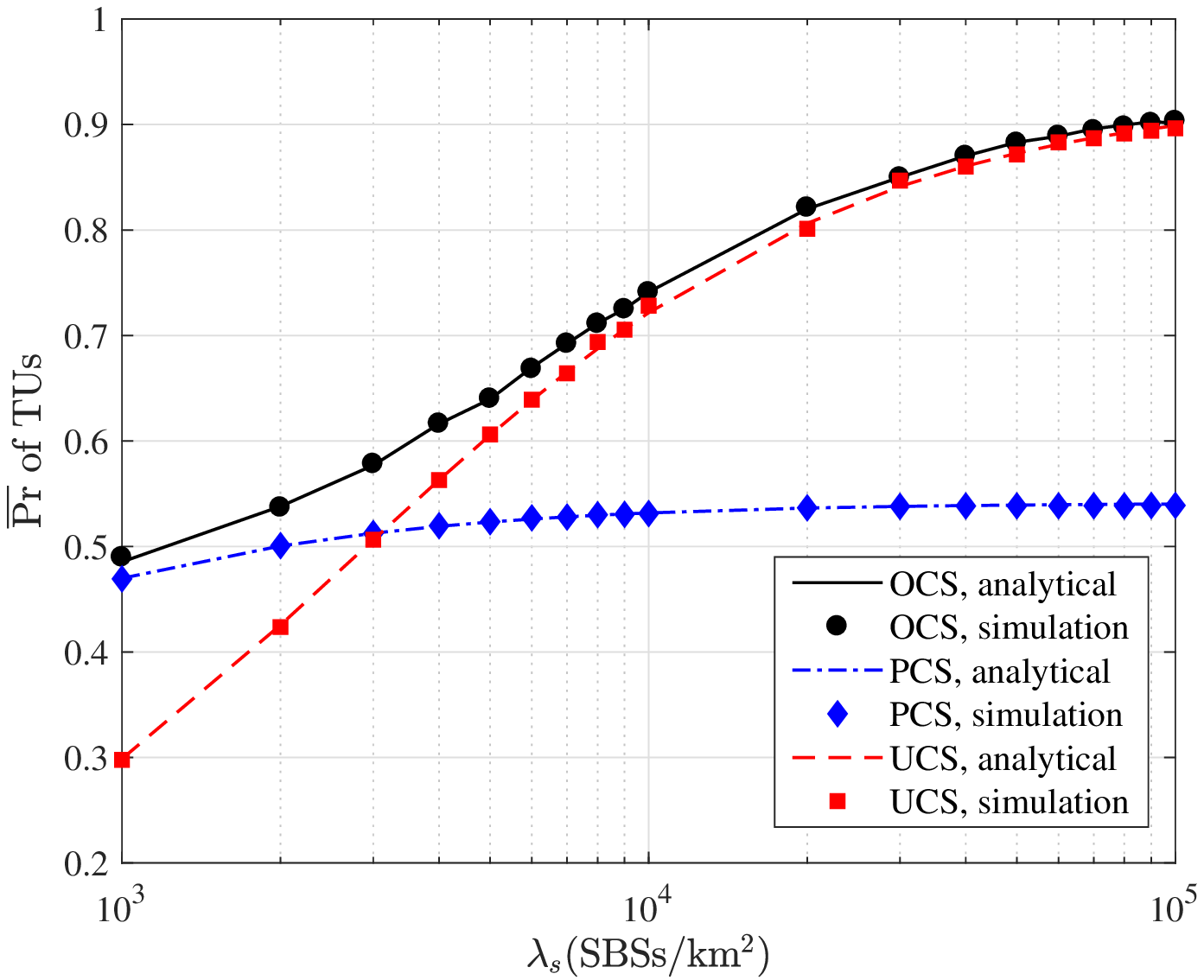}}
\subfigure[$\overline \Pr$ of UAVs]{
\includegraphics[height=0.69\columnwidth]{figure_bs_tu.eps}}
\caption{The impacts of the SBS density $\lambda_{\rm{s}}$ on the $\overline \Pr$ of TUs and the $\overline \Pr$ of UAVs with $h_{\rm{AU}}$ = 30m, $N$ = 100, $M$ = 10 and $\beta$ = 1.0.}
\end{figure}

  Fig. 2(a) shows the $\overline \Pr$ of TUs versus $\lambda_{\rm{s}}$. As for the PCS, $\overline \Pr$ increases slowly with $\lambda_{\rm{s}}$ and becomes saturated when $\lambda_{\rm{s}} > 10^4$ ${\rm{SBSs/km}^2}$. As for the UCS, $\overline \Pr$ increases more rapidly than the PCS as $\lambda_{\rm{s}}$ goes up and becomes saturated when $\lambda_{\rm{s}} > 8 \times 10^4$ ${\rm{SBSs/km}^2}$. In addition, the PCS outperforms the UCS when $\lambda_{\rm{s}} < 3\times10^3$ ${\rm{SBSs/km}^2}$, and the USC takes the lead when $\lambda_{\rm{s}} > 3\times10^3$ ${\rm{SBSs/km}^2}$.
  The performance of the PCS is comparable to that of the OCS only when $\lambda_{\rm{s}} < 10^3$ ${\rm{SBSs/km}^2}$. The same observation applies to the UCS when $\lambda_{\rm{s}} > 6\times10^4$ ${\rm{SBSs/km}^2}$.

  Fig. 2(b) shows the $\overline \Pr$ of UAVs versus $\lambda_{\rm{s}}$. This figure exhibits the similar observations to Fig. 2(a). As shown in Fig. 2(b), the PCS outperforms the UCS when $\lambda_{\rm{s}} < 1.4\times10^4$ ${\rm{SBSs/km}^2}$, and the USC takes the lead when $\lambda_{\rm{s}} > 1.4\times10^4$ ${\rm{SBSs/km}^2}$.
  The performance of the PCS is comparable to that of the OCS only when $\lambda_{\rm{s}} < 2 \times 10^3$ ${\rm{SBSs/km}^2}$. The same observation applies to the UCS when $\lambda_{\rm{s}} > 6\times10^4$ ${\rm{SBSs/km}^2}$.
  As compared to Fig. 2(a), it is observed that the average SDP of UAVs is shown to be worse than that of TUs. This is because the path loss of UAVs is severer than that of TUs according to (\ref{equ:7}), (\ref{equ:15}) and Table II.

\vspace{-3mm}
\subsection{Impact of Cache Size}
\begin{figure}[t]
 \centering
    \includegraphics[height=0.69\columnwidth]{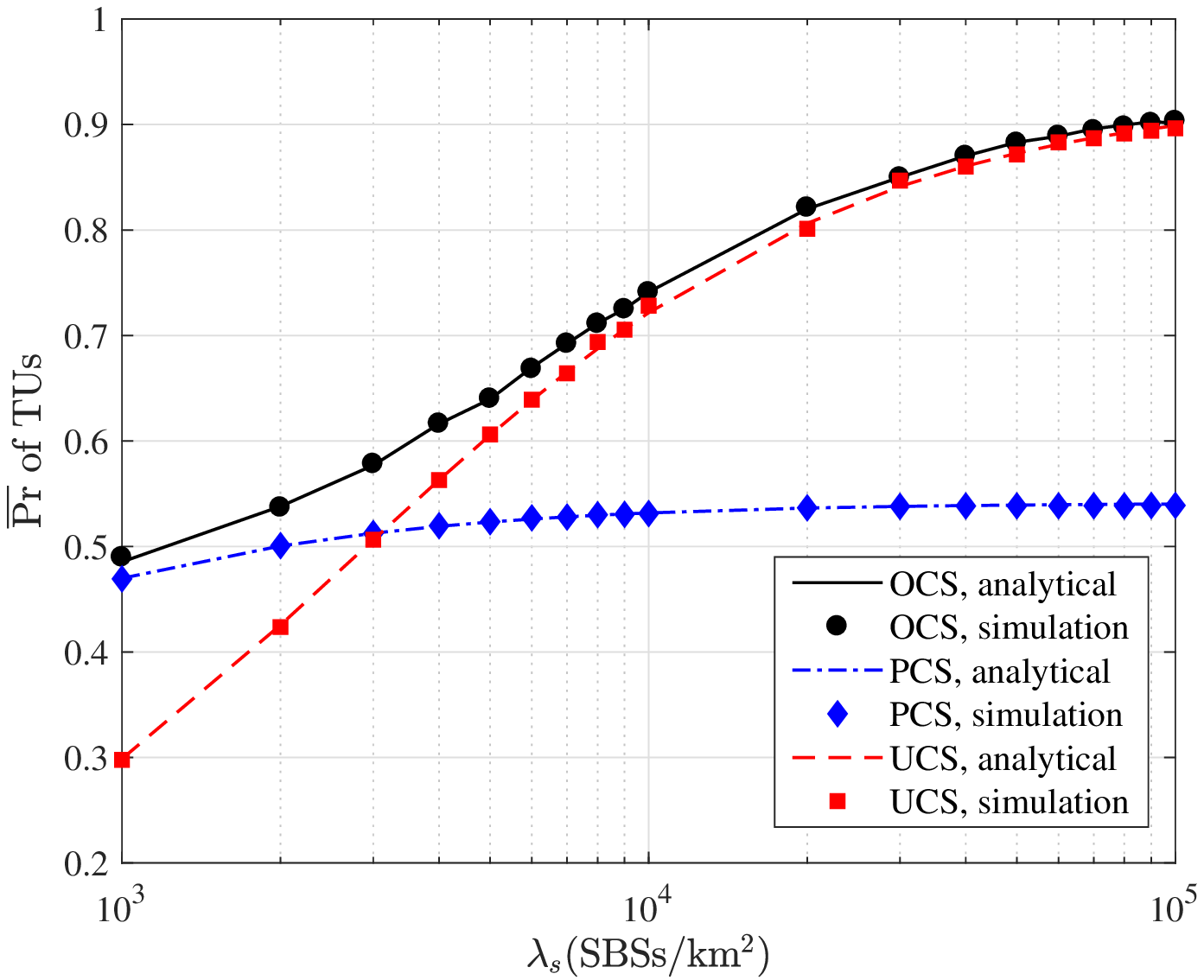}
    \caption{The impacts of the cache size $M$ on the average SDP with $\lambda_{\rm{s}}$ = $10^4$ ${\rm{SBSs/km}^2}$, $h_{\rm{AU}}$ =30m, $N$ = 100 and $\beta$ = 1.0.}
\end{figure}
  Fig. 3 compares the average SDPs $\overline {\Pr}$ among the three strategies with the cache size $M$ for TUs and UAVs respectively. Let $\lambda_{\rm{s}}$ = $10^4$ ${\rm{SBSs/km}^2}$. First, we can see that the OCS exhibits a better average SDP for TUs and UAVs than both the UCS and PCS. Second, it is observed that $\overline \Pr$ increases monotonically with the cache size. When $M = 100$, the three strategies reach the same performance. These results are consistent with $Remark$ 3. 
\begin{figure}[t]
\centering
\subfigure[$\overline \Pr$  of TUs.]{
\includegraphics[height=0.69\columnwidth]{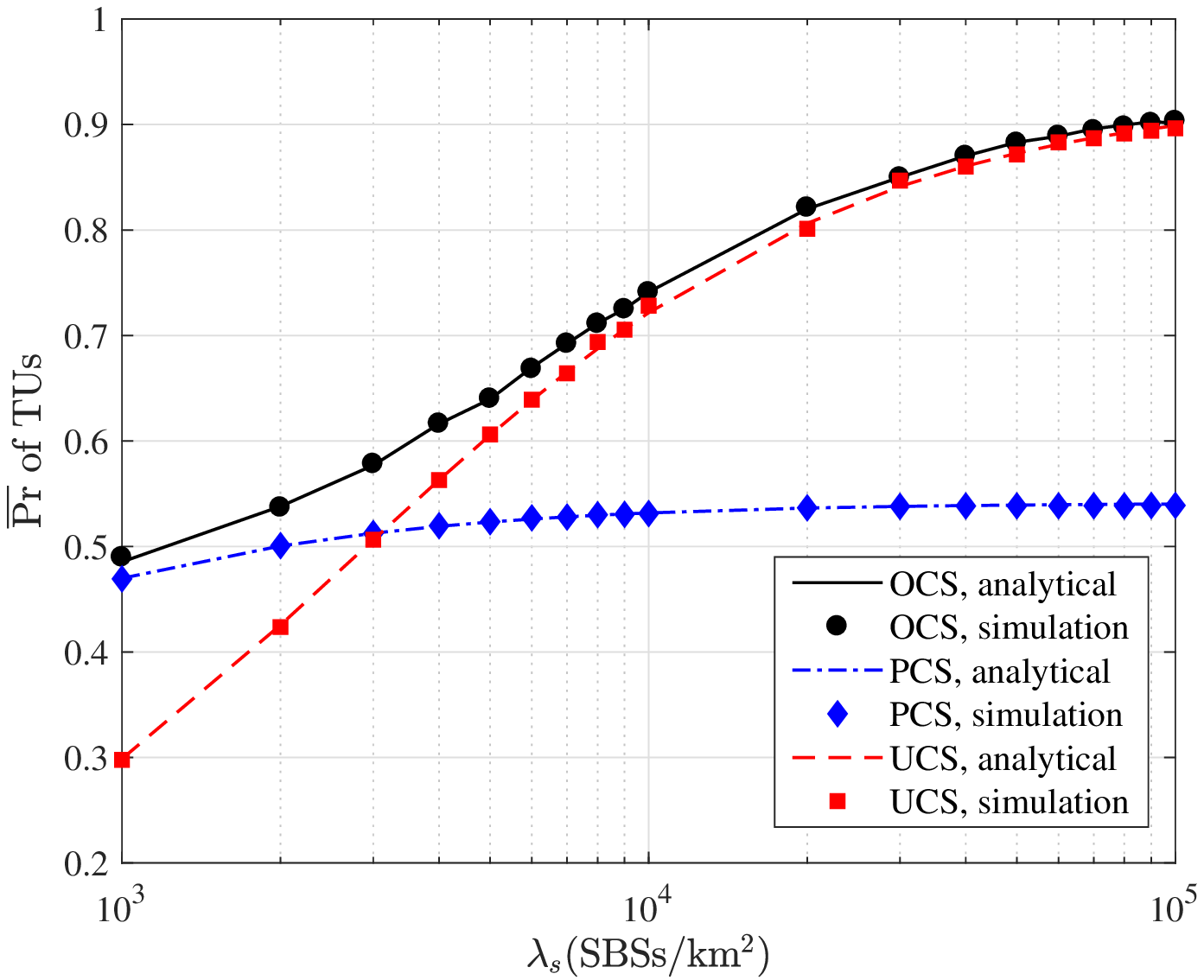}}
\subfigure[$\overline \Pr$ of UAVs.]{
\includegraphics[height=0.69\columnwidth]{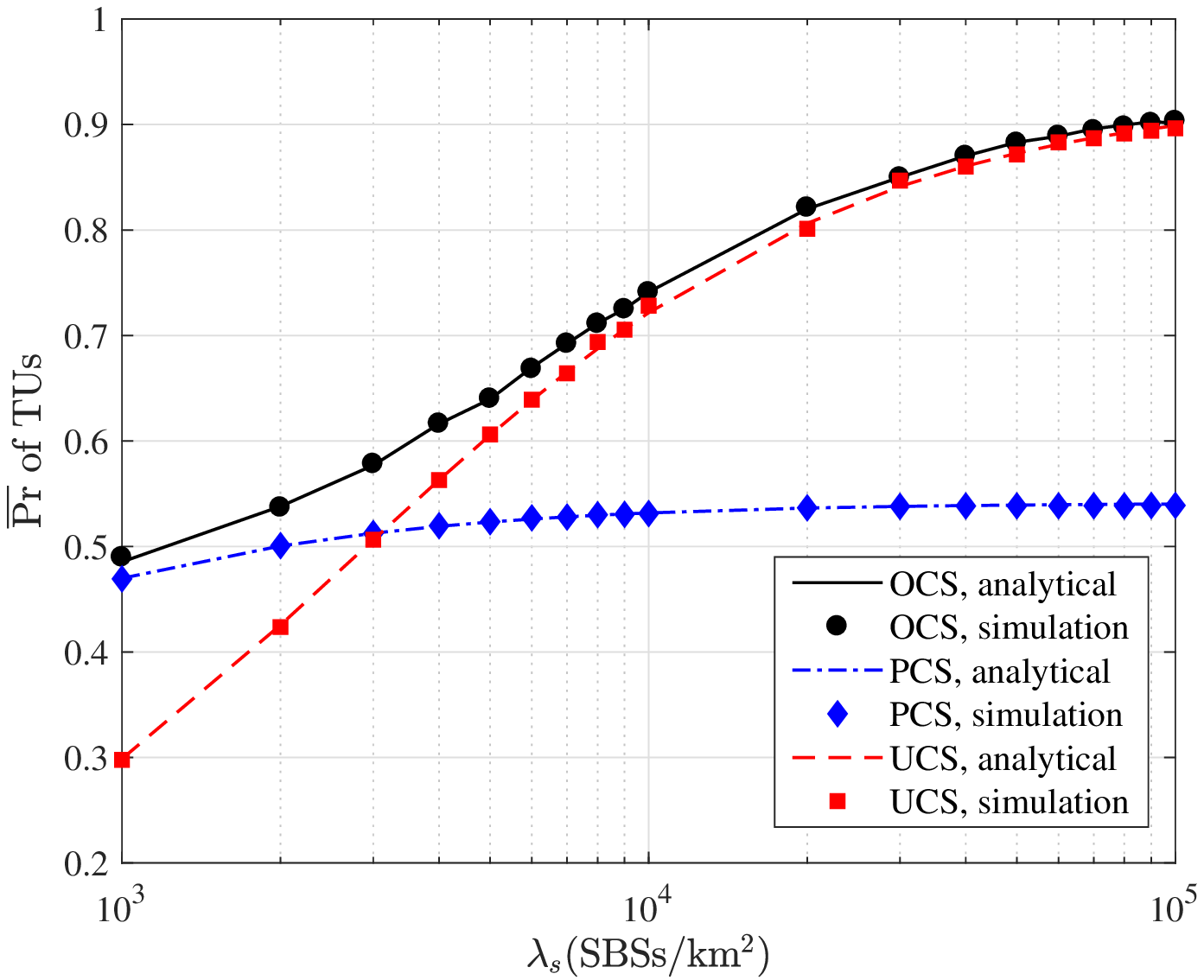}}
\caption{The impacts of the SBS density on the $\overline \Pr$ of TUs and the $\overline \Pr$ of UAVs with different cache sizes and $h_{\rm{AU}}$ =30m, $N$ = 100 and $\beta$ = 1.0.}
\end{figure}

 Fig. 4 shows the average SDPs $\overline{\Pr}$ versus the SBS density $\lambda_{\rm{s}}$ with various cache sizes for TUs and UAVs, respectively. First, it can be seen that the OCS always outperforms both the UCS and PCS. Second, $\overline \Pr$ of the PCS reaches the limit when $\lambda_{\rm{s}} \geq 10^4$ ${\rm{SBSs/km}^2}$, and the performance limit becomes larger with the increase of $M$.
 For TUs, the performance limit starts with 0.42 at $M=5$ and goes up to 0.61 at $M=15$. For UAVs, the performance limit increases from 0.37 at $M=5$ to 0.50 at $M=15$.
 Third, the $\overline \Pr$ of the UCS reaches the limit when $\lambda_{\rm{s}} \geq 5 \times 10^5$ ${\rm{SBSs/km}^2}$, the $\overline{\Pr}$ of the UCS for TUs and UAVs keeps invariant as $M$ increases when $\lambda_{\rm{s}} \geq 5 \times 10^5$ ${\rm{SBSs/km}^2}$. These observations are line with $Remark$ 2 and $Remark$ 3. For TUs and UAVs, the crossover point with the PCS and UCS achieving the same $\overline{\Pr}$ shifts to the left as $M$ increases. This is because an increase in the $M$ will boost the average SDP given a fixed $\lambda_{\rm{s}}$.

\vspace{-3mm}
\subsection{Impact of File Popularity Distribution}
\begin{figure}[t]
 \centering
    \includegraphics[height=0.69\columnwidth]{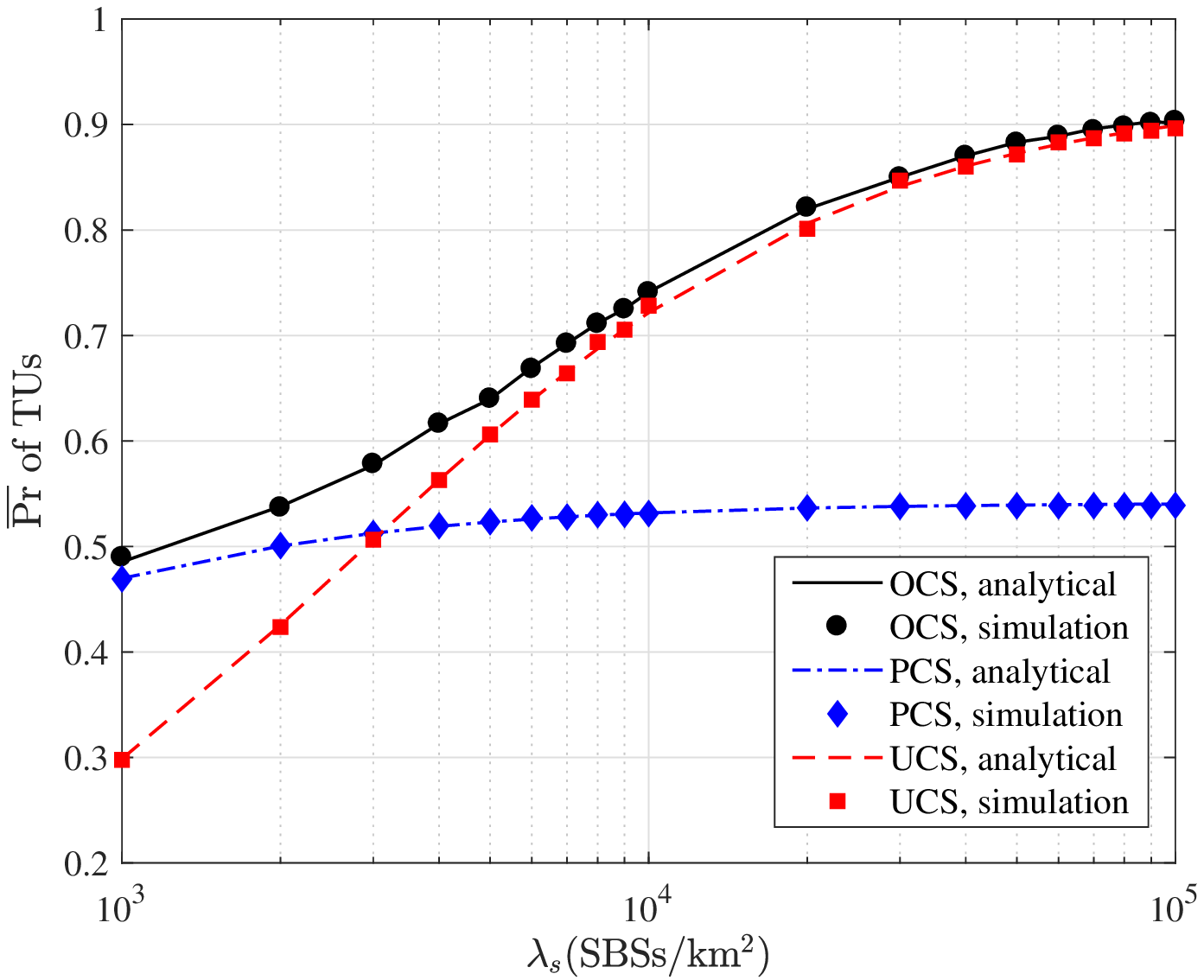}
    \caption{ The impacts of the exponent of Zipf distribution $\beta$ on the average SDP with $\lambda_{\rm{s}} = 10^4$ ${\rm{SBSs/km}^2}$, $h_{\rm{AU}}$ =30m, $N$ = 100 and $M$ = 10.}
\end{figure}
  Fig. 5 compares the average SDPs versus the exponent of Zipf distribution $\beta$ among the three strategies for TUs and UAVs respectively. We set $\lambda_{\rm{s}}$ = $10^4$ ${\rm{SBSs/km}^2}$. First, it can be seen that $\overline \Pr$ increases as $\beta$ increases. Second, the average SDP of the UCS is independent of $\beta$, since it caches each file with equal probability in the always-on architecture \cite{Caching_NTier}. However, in the dynamic on-off architecture, the performance of the UCS is slowly growing with $\beta$. According to (\ref{equ:22}) and (\ref{equ:23}), the change of $\beta$ leads to the change of $Q_n$ and eventually causes the change of the average SDP. When the $\lambda_{\rm{s}}$ is large enough, the changes of $\beta$ will not affect $\overline \Pr$ of the UCS.
  Third, it can be seen that the PCS is worse than the UCS when $\beta < 1.4$ for TUs and $\beta < 0.95$ for UAVs. As $\beta$ gradually grows, the PCS becomes better. The reason is that a few files dominate the requests and caching such popular files gives a large $\overline{\rm{Pr}}$ as $\beta$ becomes larger, since the request probabilities of files are more unevenly distributed. The average SDP of the PCS grows more rapidly with increasing $\beta$, and the average SDP of the PCS is better than that of the UCS when $\beta$ is large enough. These observations agree with $Remark$ 4.
\begin{figure}[t]
\centering
\subfigure[$\overline \Pr$  of TUs.]{
\includegraphics[height=0.69\columnwidth]{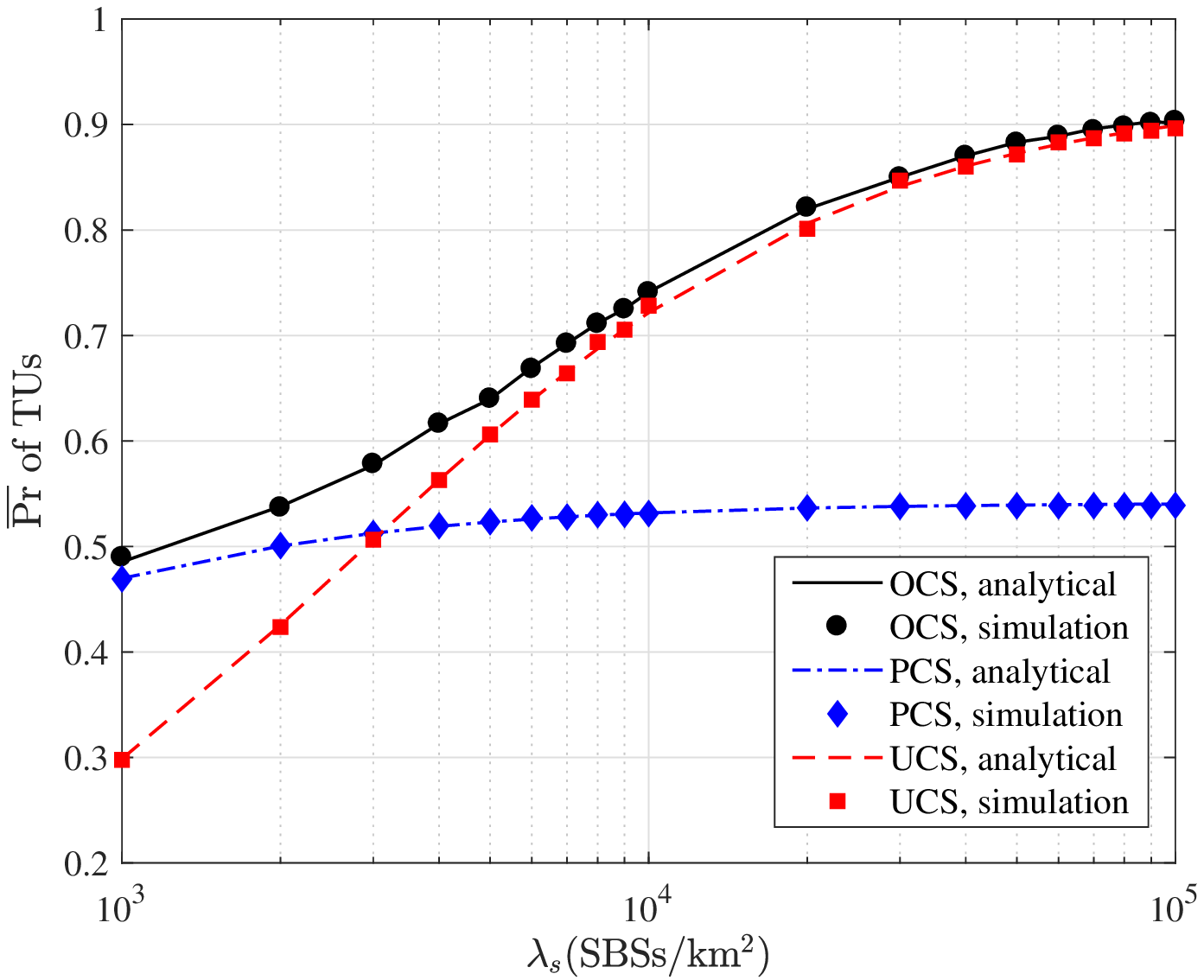}}
\subfigure[$\overline \Pr$  of UAVs.]{
\includegraphics[height=0.69\columnwidth]{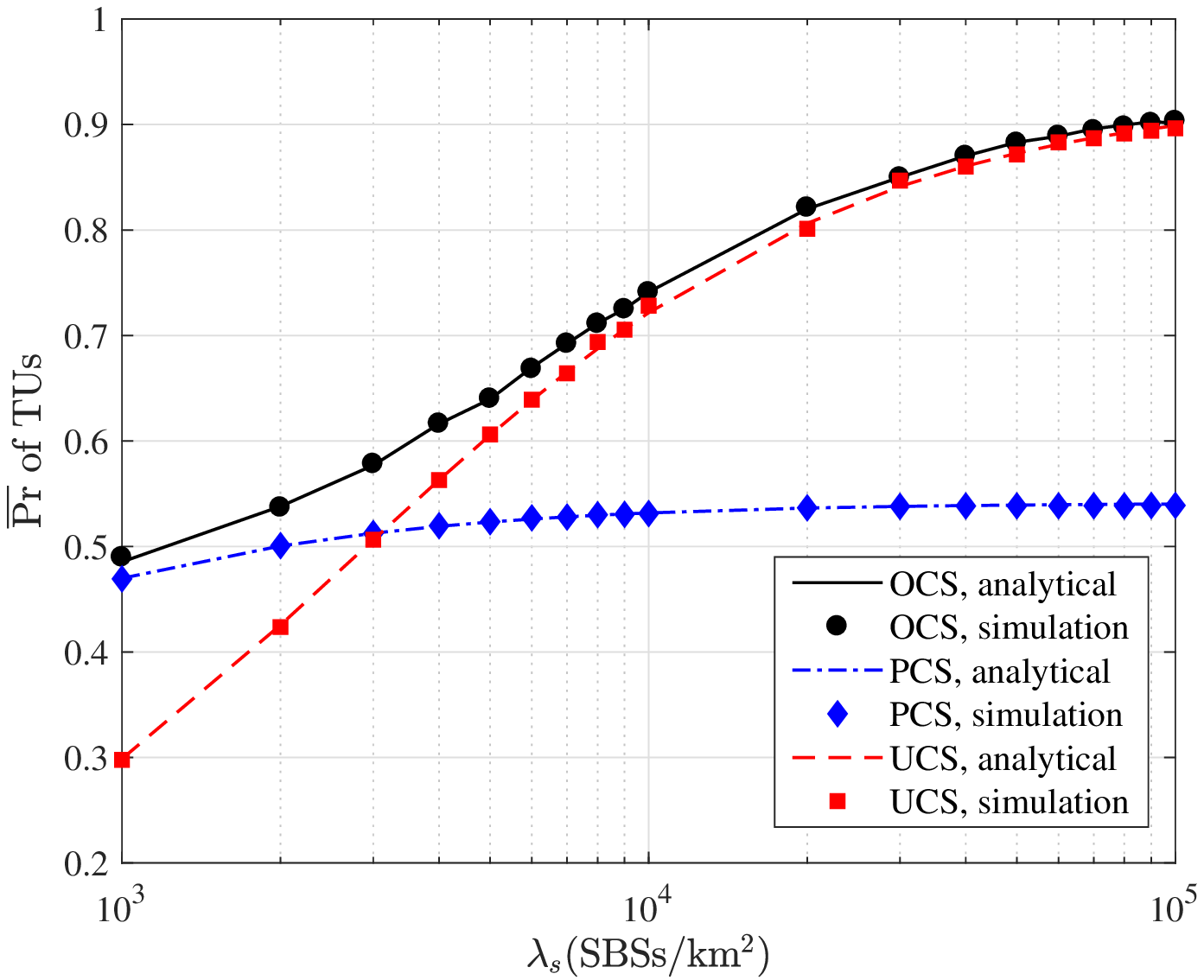}}
\caption{The impacts of the SBS density on the $\overline \Pr$ of TUs and the $\overline \Pr$ of UAVs with different $\beta$ and $\lambda_{\rm{s}}$ with $h_{\rm{AU}}$ =30m, $N$ = 100 and $M$ = 10.}
\end{figure}

  Fig. 6 shows the average SDPs versus $\lambda_{\rm{s}}$ and $\beta$ for TUs and UAVs, respectively. First, the results with a fixed $\beta$ or a fixed $\lambda_{\rm{s}}$ are consistent with that in Fig. 2 and Fig. 5 respectively. Second, as $\lambda_{\rm{s}}$ increases, we need to increase value of $\beta$ to meet the same performance of both the PCS and UCS. For example, in Fig. 6(a), the value of $\beta$ is 1.5 when $\lambda_{\rm{s}} = 10^4$ and it becomes 2.2 when $\lambda_{\rm{s}} = 10^5$. In Fig. 6(b), the value of $\beta$ is 0.9 when $\lambda_{\rm{s}} = 10^4$ and it becomes 1.5 when $\lambda_{\rm{s}} = 10^5$.

\vspace{-3mm}
\subsection{Impact of UAV Height}
\begin{figure}[t]
 \centering
    \includegraphics[height=0.69\columnwidth]{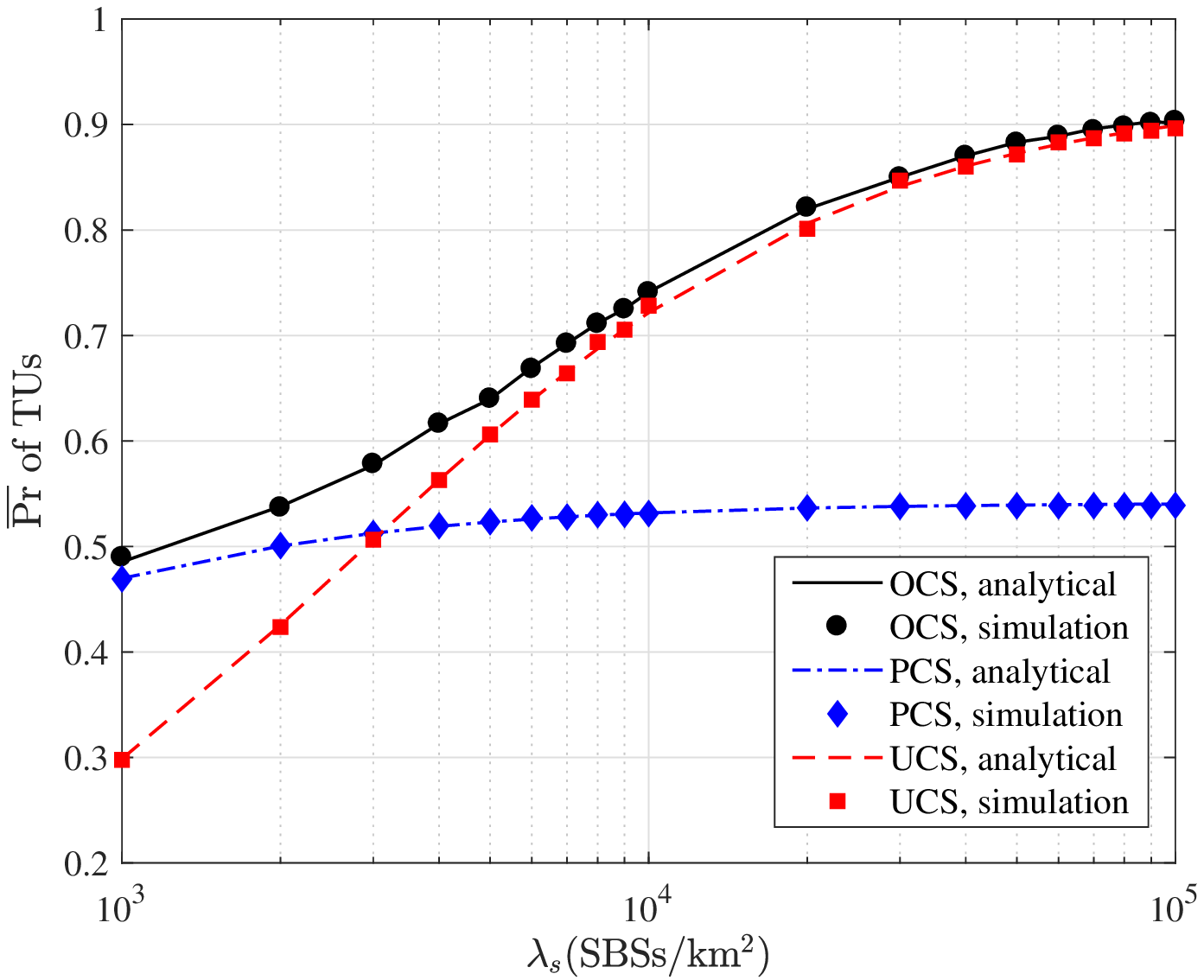}
    \caption{ The impacts of the UAV height $h_{\rm{AU}}$ on the $\overline \Pr$ of TUs/UAVs with $N$ = 100, $M$ = 10 and $\beta$ = 1.0.}
\end{figure}
 Fig. 7 depicts the impacts of the UAV height on the $\overline \Pr$ of TUs and the $\overline \Pr$ of UAVs among the three strategies. Consider that $\lambda_{\rm{s}} = 10^4$ ${\rm{SBSs/km}^2}$ and $\lambda_{\rm{s}} = 10^5$ ${\rm{SBSs/km}^2}$ respectively. First, we observe that the change of the UAV height affects the $\overline \Pr$ of UAVs but has no performance impact on TUs.
 Second, we can see that $\overline \Pr$ of UAVs decreases monotonically as the height of UAVs increases, which is consistent with $Remark$ 5. Third, similar to Fig. 2, the average SDP of the PCS almost remains the same when $\lambda_{\rm{s}}$ varies from $10^4$ ${\rm{SBSs/km}^2}$ to $10^5$ ${\rm{SBSs/km}^2}$, while the average SDP of the UCS increases over the same range of $\lambda_s$. When $\lambda_{\rm{s}} = 10^4$ ${\rm{SBSs/km}^2}$, the PCS is better than the UCS. When $h_{\rm{AU}} < 50\rm{m}$, the UCS is better than the PCS when $\lambda_{\rm{s}} = 10^5$ ${\rm{SBSs/km}}^2$. However, the PCS is better than the UCS when $h_{\rm{AU}} > 50\rm{m}$. This is because the path loss of UAV is generally large such that $\lambda_{\rm{s}}$ is not dense enough to support the average SDP of the UCS.
  \vspace{-0.1cm}
  \section{Conclusion}
  In this paper, we have developed an optimized probabilistic small-cell caching strategy for small-cell networks with TUs and UAVs to maximize the average SDP. Our analytical results have shown that the OCS can achieve a better average SDP than the PCS and UCS. Moreover, we have further analyzed the impacts of key parameters on the average SDP of TUs and UAVs and obtained the following valuable insights verified by the extensive simulation results:
  \begin{enumerate}
    \item The average SDP increases as either of the SBS density, the cache size, or the exponent of Zipf distribution increases. With the increase of the UAV height, the average SDP of UAVs decreases.
    \item When the SBS density $\lambda_{\rm{s}}$ is relatively small, the PCS achieves a better average SDP than the UCS. As the density increases, the performance of the UCS gradually improves and outperforms that of the PCS.
    \item When the exponent of Zipf distribution $\beta$ is relatively small, the UCS outperforms the PCS. As $\beta$ increases, the performance of the PCS gradually improves and surpasses that of UCS.
    \item When the cache size $M$ is equal to the popular files $N$, the average SDP of the PCS, UCS, and OCS converges, since each SBS caches all $N$ files with probability of 1.
  \end{enumerate}

  Going forward, several directions deserve further investigation. First, the uplink performance of the UAVs in the proposed caching network is yet to be analyzed. As shown in this paper, establishing the analytical results of SBSs as interferers is already non-trivial for performance analysis, the introduction of terrestrial users and UAVs as interferers in the uplink will make the analysis even more challenging.
  Second, it is of interest to consider that each SBS uses the coded caching strategy to further enhance the performance of the OCS by exploring the advantages of prefetching coded files over the uncoded placement in this paper.

  \begin{appendices}

  \section{ Proof of Theorem 1 }
  In order to evaluate ${\Pr} \left( {{D_n}} \right)$ in Theorem 1, the first key step is to calculate the PDFs for the events that the typical UE is associated with an SBS under an LoS path or a NLoS path, and the second key step is to calculate ${\Pr } \left( {{\rm{SINR}} > \delta }\right)$ for the LoS and NLoS cases conditioned on distance $r$.
  Given the piece-wise path loss model presented in (\ref{equ:1}), we have
  \begin{align}
{\Pr} \left( {{D_n}} \right) &= \int_0^\infty  {\Pr } \left( {{\rm{SINR}} > \delta } \right)f\left( r,h \right){\rm{d}}r \nonumber\\
& = \int_0^\infty  {\Pr \left[ {\textstyle{{{P\zeta \left( {r,h} \right)g}}\over{{{\sigma ^{\rm{2}}} + {I_Z}}}} > \delta } \right]} f\left( {r,h} \right){\rm{d}}r \nonumber\\
&=  {\int_{0}^{{d_1}(h)} {\Pr \left[ {{{\textstyle{{P\zeta _1^{\rm{L}}\left( r,h \right)g} \over {{\sigma ^{\rm{2}}} + {I_Z}}}}} > \delta } \right]} f_1^{\rm{L}}\left( r,h \right){\rm{d}}r} \nonumber\\
&+  {\int_{0}^{{d_1}(h)} {\Pr \left[ {{{\textstyle{{P\zeta _1^{\rm{NL}}\left( r,h \right)g} \over {{\sigma ^{\rm{2}}} + {I_Z}}}}} > \delta } \right]} f_1^{{\rm{NL}}}\left( r,h \right){\rm{d}}r} \nonumber \\
&+  \cdots \nonumber \\
&+  {\int_{{d_{K - 1}(h)}}^{\infty} {\Pr \left[ {{{\textstyle{{P\zeta _K^{\rm{L}}\left( r,h \right)g} \over {{\sigma ^{\rm{2}}} + {I_Z}}}}} > \delta } \right]} f_K^{\rm{L}}\left( r,h \right){\rm{d}}r} \nonumber\\
&+  {\int_{{d_{K - 1}(h)}}^{\infty} {\Pr \left[ {{{\textstyle{{P\zeta _K^{\rm{NL}}\left( r,h \right)g} \over {{\sigma ^{\rm{2}}} + {I_Z}}}}} > \delta } \right]} f_K^{{\rm{NL}}}\left( r,h \right){\rm{d}}r} \nonumber \\
&  = \sum\limits_{k = 1}^K {\left( {\int_{{d_{k - 1}}(h)}^{{d_k}(h)} {\Pr \left[ {{\textstyle{{P\zeta _k^{\rm{L}}\left( {r,h} \right)g} \over {{\sigma ^{\rm{2}}} + {I_Z}}}}\! >\! \delta } \right]} f_k^{\rm{L}}\left( {r,h} \right){\rm{d}}r} \right.} \nonumber \\
&\left. { + \int_{{d_{k - 1}}(h)}^{{d_k}(h)} {\Pr \!\left[\! {{\textstyle{{P\zeta _k^{{\rm{NL}}}\left( {r,h} \right)g} \over {{\sigma ^{\rm{2}}} + {I_Z}}}}\! >\! \delta } \right]} f_k^{{\rm{NL}}}\!\left(\! {r,h} \right){\rm{d}}r} \!\right)\!,
\end{align}
where $f_k^{\rm{L}}\left( r,h \right)$ and $f_k^{{\rm{NL}}}\left( r,h \right)$ are the PDFs of LoS path and NLoS path respectively. Moreover, let  $T_k^{\rm{L}}=\int_{{d_{k - 1}(h)}}^{{d_{k}(h)}} {\Pr \left[ {{{\textstyle{{P\zeta _k^{\rm{L}}\left( r,h \right)g} \over {{\sigma ^{\rm{2}}} + {I_Z}}}}} > \delta } \right]}f_k^{\rm{L}}\left( r,h \right){\rm{d}}r$ and
$T_k^{\rm{NL}}\!=\!{\int_{{d_{k - 1}(h)}}^{{d_{k}(h)}} {\Pr \left[\! {{{\textstyle{{P\zeta _k^{\rm{NL}}\left( r,h \right)g} \over {{\sigma ^{\rm{2}}} + {I_Z}}}}} \!> \!\delta } \right]} f_k^{{\rm{NL}}}\left( r,h \right){\rm{d}}r}$ respectively. Therefore, we have $\Pr(D_n)= \sum\limits_{k = 1}^K {\left( {T_k^{\rm{L}} + T_k^{{\rm{NL}}}} \right)}$.

 In the following, we discuss how to obtain $f_k^{\rm{L}}\left( r,h \right)$ and $f_k^{{\rm{NL}}}\left( r,h \right)$.

 Define $B_k^{\rm{L}}$ as the event that the signal comes from the $k$-th piece LoS path. By definition, $f_k^{\rm{L}}\left( r,h \right) = {f_{{\kern 1pt} k\left| {B_k^{\rm{L}}} \right.}}\left( {r,h\left| {B_k^{\rm{L}}} \right.} \right)\Pr \left[ {B_k^{\rm{L}}} \right]$, where $\Pr \left[ {B_k^{\rm{L}}} \right] = {\Pr} _k^{\rm{L}}\left( {r,h } \right)$ according to (3) and ${f_{{\kern 1pt} k\left| {B_k^{\rm{L}}} \right.}}\left( {r,h\left| {B_k^{\rm{L}}} \right.} \right)$ jointly characterize the following independent sub-events:

  1) For the typical UE, its serving SBS $b_o$ exists with the horizontal distance $r$ from the UE, and the corresponding unconditional PDF of $r$ is $2{\pi}r{\lambda}$ \cite{Tractable_Approach}.

  2) The probability that the LoS SBS $b_o$ in event $B_k^{\rm{L}}$ has a better link to the typical UE than any other LoS SBSs is~\cite{LoS_NLoS}
  \begin{equation}
  p_k^{\rm{L}}\left( r,h \right) = \exp \left( { - \int_0^r {{{\Pr }^{\rm{L}}}\left( r,h \right)\;2\pi\lambda u } {\rm{d}}u} \right).
  \end{equation}

  3) The probability that the LoS SBS $b_o$ in event $B_k^{\rm{L}}$ has a better link to the typical UE than any other NLoS SBSs is~\cite{LoS_NLoS}
  \begin{equation}
  p_k^{{\rm{NL}}}\left( r,h \right) \!=\! \exp \left( { - \int_0^{{r_1}} {\left( {1 \!-\! {{\Pr }^{\rm{L}}}\left( r,h \right)} \right)\;2\pi\lambda u} {\rm{d}}u} \right),
  \end{equation}
  where ${r_1} = \mathop {\arg }\limits_{{r_1}} \left\{ {{\zeta ^{{\rm{NL}}}}({r_1},h) = \zeta _k^{\rm{L}}(r,h)} \right\}$.

  With reference to \cite{LoS_NLoS}, we obtain
  \begin{equation}
  {f_{k\left| {B_k^{\rm{L}}} \right.}}\left( {r,h\left| {B_k^{\rm{L}}} \right.} \right) = p_k^{{\rm{NL}}}\left( r,h \right)p_k^{\rm{L}}\left( r,h \right)2\pi r\lambda.\!
  \end{equation}

  Thus, $f_k^{\rm{L}}\left( r,h \right)$ for $n$-th tier can be written as
  \begin{align}
  f_k^{\rm{L}}\!\left( r,h \right)= &\exp \!\left( \!{ - \!\int_0^{{r_1}} {2\pi{S_n}{\lambda _{\rm{s}}}\!\left(\! {1 \!-\! {\rm{Pr}} _k^{\rm{L}}\left( u,h \right)} \!\right)\!u{\rm{d}}u} } \!\right)\!\nonumber\\
 \times &\exp \left( { - \int_0^r {2\pi {S_n}{\lambda _{\rm{s}}}{\rm{Pr}} _k^{\rm{L}}\left( u,h \right)u{\rm{d}}u} } \!\right)\! \nonumber\\
 \times & {\rm{Pr}}_k^{\rm{L}}\left( r,h \right) 2\pi r{S_n}{\lambda _{\rm{s}}},\; {{d_{k\! -\! 1}(h)} \!<\! r \!\le\! {d_k}(h)}.
  \end{align}

  In a similar way, $f_k^{\rm{NL}}\left( r,h \right)$ for $n$-th tier can be written as
  \begin{align}
f_k^{{\rm{NL}}}\!\left(\! r,h \!\right)\!=\! &\exp \left( { - \int_0^{{r_2}} {2\pi {S_n}{\lambda _{\rm{s}}}{\Pr} _k^{\rm{L}}\left( u,h \right)u{\rm{d}}u} } \right)\nonumber\\
  \!\times\! & \exp \left( { - \int_0^r {2\pi {S_n}{\lambda _{\rm{s}}} \left({1 - {\Pr} _k^{\rm{L}}\left( u,h \right)} \right)u{\rm{d}}u} } \right) \nonumber\\
  \!\times\! &\left( {1 \!-\! {\Pr} _k^{\rm{L}}\left( r,h \right)} \right) 2\pi r{S_n}{\lambda _{\rm{s}}},{{d_{k \!-\! 1}(h)} \!<\! r \!\le\! {d_k}(h)},
  \end{align}
  where ${r_2} = \mathop {\arg }\limits_{{r_2}} \left\{ {{\zeta ^{{\rm{L}}}}({r_2},h) = \zeta _k^{\rm{NL}}(r,h)} \right\}$.

  \section{ Proof of Lemma 1 }
  Given $I_Z = I_{Z1} + I_{Z2}$, we have
  \begin{align}
&{\mathscr{L} _{I_Z}}\left( {{\textstyle{\delta  \over {P\zeta_k {\rm{(}}r,h{\rm{)}}}}}} \right) = {E_{I_Z}}\left[ {\exp \left( { - {\textstyle{{\delta {I_Z}} \over {P\zeta_k (r,h)}}}} \right)} \right] \nonumber\\
&\!=\! {E_{{I_{Z1}}}}\left[ {\exp \left( { - {\textstyle{{\delta {I_{Z1}}} \over {P\zeta_k (r,h)}}}} \right)} \right] {E_{{I_{Z2}}}}\left[ {\exp \left( { - {\textstyle{{\delta {I_{Z2}}} \over {P\zeta_k (r,h)}}}} \right)} \right].
  \end{align}

  Since the distribution of the SBSs in the $i$-th tier is viewed as a thinned HPPP $\phi_i$ with density of $S_i\lambda_s$, for the interference from the $i$-th tier, we have
  \begin{small}
  \begin{align}
&{E_{{I_{Z1}}}}\left[ {\exp \left( { - {\textstyle{{\delta {I_{Z1}}} \over {P{\zeta_k}\left( r,h \right)}}}} \right)} \right]\nonumber\\
&\!=\! {E_{{g_{{u}}},{u}}}\left[ {\prod\limits_{{u} \in \sum\nolimits_{i = 1,i \ne n}^N {\;{\phi _i}} } {\exp \left( { - {\zeta_k}{{\left( r,h \right)}^{ - 1}}\delta {g_{{u}}}{\zeta}\left( {u,h } \right)} \right)} } \right] \nonumber\\
 &\!=\! \exp \!\left( {\!- \sum\limits_{i = 1,i \ne n}^N \!{ 2\pi\Pr \!\left(\! {{A_i}} \right){S_i}{\lambda _{\rm{s}}}} \!\int_0^\infty \! {\!\left(\! {1 \!-\! {\textstyle{1 \over {1 \!+\! {\zeta_k}{{\!\left(\! r,h\!\right)\!}^{ - 1}}\delta {\zeta}\!\left(\! {u},h \!\right)\!}}}} \!\right)\!} {u}{\rm{d}}{u}} \!\right)\!.\!
\end{align}
\end{small}
  For LoS or NLoS signal, 
  \begin{align}
&\int_0^\infty  {\left( {1 - {\textstyle{1 \over {1 + {\zeta_k ^{\rm{{\left\{ L,NL \right\}}}}}{{\left( r,h \right)}^{ - 1}}\delta \zeta \left( {u},h \right)}}}} \right)} u{\rm{d}}u \nonumber\\
&= \int_0^\infty  {{\textstyle{{{{\Pr }^{\rm{L}}}\left( {u },h \right)u} \over {1 + {\zeta_k ^{\rm{{\left\{ L,NL \right\}}}}}\left( r,h \right){{(\delta {\zeta ^{\rm{L}}}\left( u,h \right))}^{ - 1}}}}}} {\rm{d}}u \nonumber\\
&+ \int_0^\infty  {{\textstyle{{\left[ {1 - {{\Pr }^{\rm{L}}}\left( {u },h \right)} \right]u} \over {1 + {\zeta_k ^{\rm{{\left\{ L,NL \right\}}}}}\left( r,h \right){{(\delta {\zeta ^{\rm{NL}}}\left( u ,h \right))}^{ - 1}}}}}} {\rm{d}}u.
\end{align}

  Likewise, for the interference from the $n$-th tier, we have
  \begin{small}
  \begin{align}
&{E_{{I_2}}}\left[ {\exp \left( { - {\textstyle{{\delta {I_{Z2}}} \over {P{\zeta_k}\left( r,h \right)}}}} \right)} \right]\nonumber\\
&\!=\!\exp \!\left(\! { \!-\! 2\pi \Pr \left( {{A_n}} \right){S_n}{\lambda _{\rm{s}}}\int_r^\infty  {\!\left( \!{1\! -\! {\textstyle{1 \over {1 \!+\! {\zeta_k}{{\left( r,h \right)}^{ \!-\! 1}}\delta {\zeta}\!\left(\! {u },h \!\right)\!}}}} \right)} {u}{\rm{d}}{u}} \right).
\end{align}
\end{small}
  For LoS or NLoS signal,
  \begin{align}
&\int_r^\infty  {\left( {1 - {\textstyle{1 \over {1 + {\zeta_k ^{\rm{{\left\{ L,NL \right\}}}}}{{\left( r,h \right)}^{ - 1}}\delta \zeta \left( {u },h \right)}}}} \right)} u{\rm{d}}u \nonumber\\
&= \int_{{\left\{ r,r_2 \right\}}}^\infty  {{\textstyle{{{{\Pr }^{\rm{L}}}\left( {u },h \right)u} \over {1 + {\zeta_k ^{\rm{{\left\{ L,NL \right\}}}}}\left( r,h \right){{(\delta {\zeta ^{\rm{L}}}\left( u,h \right))}^{ - 1}}}}}} {\rm{d}}u \nonumber\\
&+ \int_{{{\left\{ r_1,r \right\}}}}^\infty  {{\textstyle{{\left[ {1 - {{\Pr }^{\rm{L}}}\left( {u },h \right)} \right]u} \over {1 + {\zeta_k ^{\rm{{\left\{ L,NL \right\}}}}}\left( r,h \right){{(\delta {\zeta ^{\rm{NL}}}\left( u,h \right))}^{ - 1}}}}}} {\rm{d}}u.
\end{align}

  \section{ Proof of Theorem 2 }

  Consider that $r\!\to\!0$ and neglecting noise. When $\lambda_{\rm{s}} \!\to\! +\! \infty $, the average SDP of TUs over all possible $N$ files is given by
  \begin{small}
  \begin{align}
&{{\Pr} _{\rm{t}}}\left( D \right) = \sum\limits_{n = 1}^N {{Q_n}} {{\Pr} _{\rm{t}}}\left( {{D_n}} \right) \nonumber\\
&= \sum\limits_{n = 1}^N {{Q_n}\int_0^{{d_{\rm{T}}}} {{\mathscr{L} _{I_Z}}\left( {{\textstyle{{\delta {l^{\alpha _{\rm{t}}^{\rm{L}}}}} \over {P{\rm{A}}_{\rm{t}}^{\rm{L}}}}}} \right)} {f_{\rm{t}}}\left( r,h_1 \right){\rm{d}}r} \nonumber\\
 &= \sum\limits_{n = 1}^N {{Q_n}} \left\{ {\int_0^{{d_{\rm{T}}}} {2\pi {S_n}{\lambda _{\rm{s}}}r\exp \left( { - \pi {S_n}{\lambda _{\rm{s}}}{r^2}} \right)} } \right.\nonumber\\
 &\left. \times {\exp \left( {\sum\limits_{i = 1,i \ne n}^N {{Q_i}} B\left( {r,{h_1}} \right) + {Q_n}C\left( {r,{h_1}} \right)} \right){\rm{d}}r} \right\} ,
\end{align}
\end{small}
where
\begin{small}
 \begin{align}
 \label{equ:b}
 & B\left( {r,{h_1}} \right)\! =\! -\! 2\pi {\lambda _{\rm{u}}}\left[ {\int_0^{{d_{\rm{T}}}} \textstyle{ {\frac{u}{{1 + \zeta _{\rm{t}}^{\rm{L}}\left( {r,{h_1}} \right){{(\delta A_{\rm{t}}^{\rm{L}})}^{ - 1}}{{\left( {\sqrt {{u^2} + h_1^2} } \right)}^{\alpha _{\rm{t}}^{\rm{L}}}}}}} } } \right. \nonumber\\
 & \!\times\! \textstyle\left( {1 - \frac{{\sqrt {{u^2} + h_1^2} }}{{{l_0}}}} \right){\rm{d}}u  \nonumber\\
 &\!+\! \int_0^{{d_{\rm{T}}}} \textstyle{ {\frac{u}{{1 + \zeta _{\rm{t}}^{\rm{L}}\left( {r,{h_1}} \right){{(\delta A_{\rm{t}}^{{\rm{NL}}})}^{ - 1}}{{\left( {\sqrt {{u^2} + h_1^2} } \right)}^{\alpha _{\rm{t}}^{{\rm{NL}}}}}}}} }\!\times\! \frac{{\sqrt {{u^2} + h_1^2} }}{{{l_0}}}{\rm{d}}u \nonumber\\
 &\left. { \!+\! \int_{{d_{\rm{T}}}}^\infty  \textstyle{\frac{u}{{1 \!+\! \zeta _{\rm{t}}^{\rm{L}}\left( {r,{h_1}} \right){{(\delta A_{\rm{t}}^{{\rm{NL}}})}^{ - 1}}{{\left( {\sqrt {{u^2} + h_1^2} } \right)}^{\alpha _{\rm{t}}^{{\rm{NL}}}}}}}} {\rm{d}}u} \right],\\
\label{equ:c}
 & C\left( {r,{h_1}} \right) \! =\!  -\! 2\pi {\lambda _{\rm{u}}}\left[ {\int_{r}^{{d_{\rm{T}}}} {{\textstyle{u \over {1 + \zeta _{\rm{t}}^{\rm{L}}\left( {r,{h_1}} \right){{(\delta A_{\rm{t}}^{\rm{L}})}^{ - 1}}{{\left( {\sqrt {{u^2} + h_1^2} } \right)}^{\alpha _{\rm{t}}^{\rm{L}}}}}}}} } \right. \nonumber\\
 & \!\times\! \textstyle {\left( {1 - {{{\sqrt {{u^2} + h_1^2} }}\over{{{l_0}}}}} \right)}{\rm{d}}u  \nonumber\\
 &\!+\! \int_{r_1}^{{d_{\rm{T}}}} {{\textstyle{u \over {1 + \zeta _{\rm{t}}^{\rm{L}}\left( {r,{h_1}} \right){{(\delta A_{\rm{t}}^{\rm{NL}})}^{ - 1}}{{\left( {\sqrt {{u^2} + h_1^2} } \right)}^{\alpha _{\rm{t}}^{\rm{NL}}}}}}}} \!\times\! \textstyle\frac{{\sqrt {{u^2} + h_1^2} }}{{{l_0}}}{\rm{d}}u \nonumber\\
 &\left. { \!+\! \int_{{d_{\rm{T}}}}^\infty  \textstyle{\frac{u}{{1 \!+\! \zeta _{\rm{t}}^{\rm{L}}\left( {r,{h_1}} \right){{(\delta A_{\rm{t}}^{{\rm{NL}}})}^{ - 1}}{{\left( {\sqrt {{u^2} + h_1^2} } \right)}^{\alpha _{\rm{t}}^{{\rm{NL}}}}}}}} {\rm{d}}u} \right].
\end{align}
\end{small}

  Consider that $r \!\to\! 0$ and neglecting noise. When $\lambda_{\rm{s}} \!\to\!  + \!\infty $, the average SDP of UAVs over all possible $N$ files is given by
  \begin{small}
  \begin{align}
&{{\Pr} _{\rm{a}}}\left( D \right) = \sum\limits_{n = 1}^N {{Q_n}} {{\Pr} _{\rm{a}}}\left( {{D_n}} \right) \nonumber\\
 &= \sum\limits_{n = 1}^N {{Q_n}} \left\{ {\int_0^{{d_{\rm{A}}}} {2\pi {S_n}{\lambda _{\rm{s}}}r\exp \left( { - \pi {S_n}{\lambda _{\rm{s}}}{r^2}} \right)} } \right.\nonumber\\
 &\left. \times {\exp \left( {\sum\limits_{i = 1,i \ne n}^N {{Q_i}} E\left( {r,{h_2}} \right) + {Q_n}F\left( {r,{h_2}} \right)} \right){\rm{d}}r} \right\} ,
\end{align}
\end{small}
where
  \begin{small}
  \begin{align}
  \label{equ:e}
&E\left( {r,h_2} \right) \! =\! - 2\pi {\lambda _{\rm{u}}}\left[ {\int_0^{{d_{\rm{A}}}} {\left( {{\textstyle{1 \over {1 + \zeta _{\rm{a}}^{\rm{L}}\left( r,h_2 \right){{(\delta A_{\rm{a}}^{\rm{L}})}^{ - 1}}{{\left( {\sqrt {{u}^2 + {h_2^2}} } \right)}^{\alpha _{\rm{a}}^{\rm{L}}}}}}}} \right)} {u}{\rm{d}}{u}} \right.\nonumber \\
 &+ \int_{{d_{\rm{A}}}}^\infty  {\left( {{\textstyle{1 \over {1 + \zeta _{\rm{a}}^{\rm{L}}\left( r,h_2 \right){{(\delta A_{\rm{a}}^{\rm{L}})}^{ - 1}}{{\left( {\sqrt {{u}^2 + {h_2^2}} } \right)}^{\alpha _{\rm{a}}^{\rm{L}}}}}}}} \right)} \nonumber \\
  &\times \left( {{\textstyle{{{d_{\rm{A}}}} \over {{u}}}} + \exp\left( {{\textstyle{{ - {u}} \over {{p_1}}}}} \right)\left( {1 - {\textstyle{{{d_{\rm{A}}}} \over {{u}}}}} \right)} \right){u}{\rm{d}}{u} \nonumber \\
  &+ \int_{{d_{\rm{A}}}}^\infty  {\left( {{\textstyle{1 \over {1 + \zeta _{\rm{a}}^{\rm{L}}\left( r,h_2 \right){{(\delta A_{\rm{a}}^{{\rm{NL}}})}^{ - 1}}{{\left( {\sqrt {{u}^2 + {h_2^2}} } \right)}^{\alpha _{\rm{a}}^{{\rm{NL}}}}}}}}} \right)} \nonumber \\
 &\left. { \times \left( {1 - {\textstyle{{{d_{\rm{A}}}} \over {{u}}}} - \exp\left( {{\textstyle{{ - {u}} \over {{p_1}}}}} \right)\left( {1 - {\textstyle{{{d_{\rm{A}}}} \over {{u}}}}} \right)} \right){u}{\rm{d}}{u}} \right],
\end{align}
\end{small}
\begin{small}
  \begin{align}
  \label{equ:f}
&F\left( {r,h_2} \right) \! =\! - 2\pi {\lambda _{\rm{u}}}\left[ {\int_r^{{d_{\rm{A}}}} {\left( {{\textstyle{1 \over {1 + \zeta _{\rm{a}}^{\rm{L}}\left( r,h_2 \right){{(\delta A_{\rm{a}}^{\rm{L}})}^{ - 1}}{{\left( {\sqrt {{u}^2 + {h_2^2}} } \right)}^{\alpha _{\rm{a}}^{\rm{L}}}}}}}} \right)} {u}{\rm{d}}{u}} \right.\nonumber \\
 &+ \int_{{d_{\rm{A}}}}^\infty  {\left( {{\textstyle{1 \over {1 + \zeta _{\rm{a}}^{\rm{L}}\left( r,h_2 \right){{(\delta A_{\rm{a}}^{\rm{L}})}^{ - 1}}{{\left( {\sqrt {{u}^2 + {h_2^2}} } \right)}^{\alpha _{\rm{a}}^{\rm{L}}}}}}}} \right)} \nonumber \\
  &\times \left( {{\textstyle{{{d_{\rm{A}}}} \over {{u}}}} + \exp\left( {{\textstyle{{ - {u}} \over {{p_1}}}}} \right)\left( {1 - {\textstyle{{{d_{\rm{A}}}} \over {{u}}}}} \right)} \right){u}{\rm{d}}{u}\nonumber \\
  &+ \int_{{d_{\rm{A}}}}^\infty  {\left( {{\textstyle{1 \over {1 + \zeta _{\rm{a}}^{\rm{L}}\left( r,h_2 \right){{(\delta A_{\rm{a}}^{{\rm{NL}}})}^{ - 1}}{{\left( {\sqrt {{u}^2 + {h_2^2}} } \right)}^{\alpha _{\rm{a}}^{{\rm{NL}}}}}}}}} \right)} \nonumber \\
 &\left. { \times \left( {1 - {\textstyle{{{d_{\rm{A}}}} \over {{u}}}} - \exp\left( {{\textstyle{{ - {u}} \over {{p_1}}}}} \right)\left( {1 - {\textstyle{{{d_{\rm{A}}}} \over {{u}}}}} \right)} \right){u}{\rm{d}}{u}} \right].
\end{align}
\end{small}

  Overall, the average SDP of TUs and UAVs is
  \begin{small}
  \begin{align}
\overline {\Pr} & = \sum\limits_{n = 1}^N {{Q_n}} \left( {{\textstyle{{{\lambda _{\rm{TU}}}} \over {{\lambda _{\rm{u}}}}}}{{\Pr }_{\rm{t}}}\left( {{D_n}} \right) + {\textstyle{{{\lambda _{\rm{AU}}}} \over {{\lambda _{\rm{u}}}}}}{{\Pr} _{\rm{a}}}\left( {{D_n}} \right)} \right) \nonumber\\
&= \sum\limits_{n = 1}^N {{Q_n}{S_n}\left[ {\int_0^{{d_{\rm{T}}}} {{G_n}\exp \left( { - \pi {S_n}{\lambda _{\rm{s}}}{r^2}} \right)} {\rm{d}}r} \right.} \nonumber\\
&\left. { + \int_0^{{d_{\rm{A}}}} {{H_n}\exp \left( { - \pi {S_n}{\lambda _{\rm{s}}}{r^2}} \right)} {\rm{d}}r} \right],
\end{align}
\end{small}
where
  \begin{small}
  \begin{align}
  {G_n}\!=\! {\textstyle{{{\lambda _{\rm{TU}}}2\pi {\lambda _{\rm{s}}}r} \over {{\lambda _{\rm{u}}}}}} \textstyle\exp \!\left(\! {\sum\limits_{i \!=\! 1,i \!\ne\! n}^N \!{{Q_i}} B\!\left(\! {r,h_1} \!\right)\! +\! {Q_n}C\left( {r,h_1} \!\right)\!} \right)\!,\!\\
  {H_n}\! =\! {\textstyle{{{\lambda _{\rm{AU}}}2\pi {\lambda _{\rm{s}}}r} \over {{\lambda _{\rm{u}}}}}} \textstyle\exp \!\left(\! {\sum\limits_{i \!=\! 1,i \!\ne\! n}^N \!{{Q_i}} E\!\left(\! {r,h_2} \!\right) \!+\! {Q_n}F\left( {r,h_2} \!\right)\!} \right)\!.\!
  \end{align}
  \end{small}

\section{ Proof of Theorem 3 }

  According to (\ref{equ:21}), when the interference come from the $n$-th tier $I_{Z2}$, we have
\begin{align}
&\exp \left( { - 2\pi \Pr \left( {{A_n}} \right){S_n}{\lambda _{\rm{s}}}\int_r^\infty  {{\textstyle{u \over {1 + {l^{-\alpha} }{\delta ^{ - 1}}{{\left( {\sqrt {{u^2} + {h^2}} } \right)}^\alpha }}}}} {\rm{d}}u} \right) \nonumber\\
 &\!=\! \exp \left( { - \pi \Pr \left( {{A_n}} \right){S_n}{\lambda _{\rm{s}}}{\delta ^{^{{\textstyle{2 \over \alpha }}}}}{l^2}\frac{2}{\alpha }\int_{{\delta ^{ - 1}}}^\infty  {\frac{{{z^{^{{\textstyle{2 \over \alpha }}} - 1}}}}{{1 + z}}} {\rm{d}}z} \right) \nonumber\\
 &\!=\! \exp \left( {\!-\! \pi \Pr \left( {{A_n}} \right){S_n}{\lambda _{\rm{s}}}{l^2}\frac{{2\delta }}{{\alpha  \!-\! 2}}{}_2{F_1}\!\left(\! {1,1 \!-\! \frac{2}{\alpha };2 \!-\! \frac{2}{\alpha }; \!-\! \delta } \!\right)\!} \right),
\end{align}
where $z = {\delta ^{ - 1}}{l^{ - \alpha }}{\left( {\sqrt {{u^2} + {h^2}} } \right)^\alpha }$ and ${}_2{F_1}\left(  \cdot  \right)$ denotes the hyper-geometric function.

  When the interference come from other tiers $I_{Z1}$, we have
  \begin{align}
&\textstyle{\exp \left( { - 2\pi \sum\limits_{i = 1,i \ne n}^N {\Pr \left( {{A_i}} \right){S_i}{\lambda _{\rm{s}}}} \int_0^\infty  {{\textstyle{u \over {1 + {l^{-\alpha} }{\delta ^{ - 1}}{{\left( {\sqrt {{u^2} + {h^2}} } \right)}^\alpha }}}}} {\rm{d}}u} \right)}\nonumber\\
 &= \textstyle{\exp \!\left(\! {\! -\! \pi \!\sum\limits_{i = 1,i \ne n}^N \!{\Pr \!\left(\! {{A_i}} \!\right)\!{S_i}{\lambda _{\rm{s}}}} {l^2}\frac{{2\delta }}{{\alpha  - 2}} }  { {}_2{F_1}\left( {1, 1 \!-\!\frac{2}{\alpha }; 2\! -\! \frac{2}{\alpha };  \!-\! \frac{{\delta {l^{ \alpha }}}}{{{h^\alpha }}}} \!\right)\!} \!\right)\!}.
\end{align}

  Overall, we have
\begin{align}
\label{equ:24}
&\overline {\Pr} = \textstyle{\sum\limits_{n = 1}^N {{Q_n}\int_0^\infty {\pi {S_n}{\lambda _{\rm{s}}}\exp \left( { - \pi {\lambda _{\rm{s}}}{r^2}{S_n}} \right.} }} \nonumber\\
 &\textstyle{\!-\! \pi \sum\limits_{i = 1,i \ne n}^N {{\Pr} \left( {{A_i}} \right){S_i}{\lambda _{\rm{s}}}} {l^2}\frac{{2\delta }}{{\alpha  \!-\! 2}}{}_2{F_1}\left( {1,1 \!-\! \frac{2}{\alpha };2 \!-\! \frac{2}{\alpha }; \!-\! \frac{{\delta {l^{ - \alpha }}}}{{{h^\alpha }}}} \right)}\nonumber\\
&\textstyle{\left. {- \pi {\Pr} \left( {{A_n}} \right){S_n}{\lambda _{\rm{s}}}{l^2}\frac{{2\delta }}{{\alpha  - 2}}{}_2{F_1}\left( {1,1 \!-\! \frac{2}{\alpha };2 \!-\! \frac{2}{\alpha }; \!-\! \delta } \right)} \right){\rm{d}}{r^2}}.
\end{align}

When the SBS density is large enough, we rewrite (\ref{equ:24}) as
\begin{align}
\label{equ:26}
&\overline {\Pr} = \textstyle{\sum\limits_{n = 1}^N {{Q_n}\int_0^\infty  {\pi {S_n}{\lambda _{\rm{s}}}\exp \left( { - \pi {\lambda _{\rm{s}}}{r^2}{S_n}} \right.} }} \nonumber\\
&\textstyle{\left. { \!-\! \pi \sum\limits_{i \!=\! 1}^N {{\Pr} \left( {{A_i}} \right){S_i}{\lambda _{\rm{s}}}} {l^2}\frac{{2\delta }}{{\alpha  \!-\! 2}}{}_2{F_1}\left( {1,1 \!-\! \frac{2}{\alpha };2 \!-\! \frac{2}{\alpha }; \!-\! \delta } \right)} \right){\rm{d}}{r^2}} \nonumber\\
 &\textstyle{\!=\! \sum\limits_{n = 1}^N\frac{{{Q_n}{S_n}\exp \left( { - \pi {\lambda _{\rm{s}}}{h^2}\sum\limits_{i = 1}^N {\Pr \left( {{A_i}} \right){S_i}} \mathcal{F}(\delta ,\alpha )} \right)}}{{{S_n} + \sum\limits_{i = 1}^N {\Pr \left( {{A_i}} \right){S_i}} \mathcal{F}(\delta ,\alpha )}}},
\end{align}
where $\mathcal{F}\left( {\delta ,\alpha } \right) = \frac{2\delta }{{\alpha  - 2}}{}_2{F_1}\left( {1,1 - \frac{2}{\alpha };2 - \frac{2}{\alpha }; - \delta } \right)$.

 Substituting (\ref{equ:25}) into (\ref{equ:26}), we have
 \begin{align}
\overline \Pr &= \sum\limits_{n = 1}^N \textstyle{\frac{{{Q_n}{S_n}\exp \left( { - \pi {\lambda _{\rm{s}}}{h^2}\sum\limits_{i = 1}^N {\frac{{{Q_i}{\lambda _{\rm{u}}}}}{{{S_i}{\lambda _{\rm{s}}}}}{S_i}} \mathcal{F}(\delta ,\alpha )} \right)}}{{{S_n} + \sum\limits_{i = 1}^N {\frac{{{Q_i}{\lambda _{\rm{u}}}}}{{{S_i}{\lambda _{\rm{s}}}}}{S_i}} \mathcal{F}(\delta ,\alpha )}}} \nonumber \\
&= \sum\limits_{n = 1}^N \textstyle{\frac{{{Q_n}{S_n}\exp \left( { - \pi {h^2}{\lambda _u}\mathcal{F}(\delta ,\alpha )} \right)}}{{{S_n} + \frac{{{\lambda _{\rm{u}}}}}{{{\lambda _{\rm{s}}}}}\mathcal{F}(\delta ,\alpha )}}}.
\end{align}

 For the PCS, we have
  \begin{equation}
\overline{\Pr} = \sum\limits_{n = 1}^M \textstyle{{\frac{{{Q_n}\exp \left( { - \pi {h^2}{\lambda _{\rm{u}}}\mathcal{F}(\delta ,\alpha )} \right)}}{{1 + \frac{{{\lambda _{\rm{u}}}}}{{{\lambda _{\rm{s}}}}}\mathcal{F}(\delta ,\alpha )}}}}.
  \end{equation}

  For the UCS, we have
  \begin{align}
\overline{\Pr} &= \sum\limits_{n = 1}^N \textstyle{\frac{{{Q_n}{M \over N}\exp \left( { - \pi {h^2}{\lambda _{\rm{u}}}\mathcal{F}(\delta ,\alpha )} \right)}}{{{M \over N} + \frac{{{\lambda _{\rm{u}}}}}{{{\lambda _{\rm{s}}}}}\mathcal{F}(\delta ,\alpha )}}} \nonumber\\
&= \textstyle\frac{{\exp \left( { - \pi {h^2}{\lambda _{\rm{u}}}\mathcal{F}(\delta ,\alpha )} \right)}}{{1 + {\textstyle{{N{\lambda _{\rm{u}}}} \over {M{\lambda _{\rm{s}}}}}}\mathcal{F}(\delta ,\alpha )}}.
\end{align}

\end{appendices}

\bibliographystyle{IEEEtran}
\bibliography{references}

\begin{thebibliography}{10}
\providecommand{\url}[1]{#1}
\csname url@samestyle\endcsname
\providecommand{\newblock}{\relax}
\providecommand{\bibinfo}[2]{#2}
\providecommand{\BIBentrySTDinterwordspacing}{\spaceskip=0pt\relax}
\providecommand{\BIBentryALTinterwordstretchfactor}{4}
\providecommand{\BIBentryALTinterwordspacing}{\spaceskip=\fontdimen2\font plus
\BIBentryALTinterwordstretchfactor\fontdimen3\font minus
  \fontdimen4\font\relax}
\providecommand{\BIBforeignlanguage}[2]{{%
\expandafter\ifx\csname l@#1\endcsname\relax
\typeout{** WARNING: IEEEtran.bst: No hyphenation pattern has been}%
\typeout{** loaded for the language `#1'. Using the pattern for}%
\typeout{** the default language instead.}%
\else
\language=\csname l@#1\endcsname
\fi
#2}}
\providecommand{\BIBdecl}{\relax}
\BIBdecl

\bibitem{Cisco1}
Cisco, ``Cisco visual networking index: Global mobile data traffic forecast
  update 2016-2021,'' \emph{White Paper}, Feb. 2017.

\bibitem{UD_SC_Deployments}
D.~L\'opez-P\'erez, M.~Ding, H.~Claussen, and A.~H. Jafari, ``Towards 1
  {G}bps/{UE} in cellular systems: Understanding ultra-dense small cell
  deployments,'' \emph{IEEE Commun. Surveys. Tuts.}, vol.~17, no.~4, pp.
  2078--2101, Fourthquarter 2015.

\bibitem{LTE_for_UAV}
X.~Lin, V.~Yajnanarayana, S.~D. Muruganathan, S.~Gao, H.~Asplund, H.~L.
  Maattanen, M.~Bergstrom, S.~Euler, and Y.~P.~E. Wang, ``The sky is not the
  limit: {LTE} for unmanned aerial vehicles,'' \emph{IEEE Commun. Mag.},
  vol.~56, no.~4, pp. 204--210, Apr. 2018.

\bibitem{AD1}
Z.~{Kaleem} and M.~H. {Rehmani}, ``Amateur drone monitoring: State-of-the-art
  architectures, key enabling technologies, and future research directions,''
  \emph{IEEE Wireless Commun.}, vol.~25, no.~2, pp. 150--159, Apr. 2018.

\bibitem{Trajectory_Communication}
Q.~Wu, Y.~Zeng, and R.~Zhang, ``Joint trajectory and communication design for
  multi-{UAV} enabled wireless networks,'' \emph{IEEE Trans. Wireless Commun.},
  vol.~17, no.~3, pp. 2109--2121, Mar. 2018.

\bibitem{AD2}
Z.~{Kaleem}, N.~N. {Qadri}, T.~Q. {Duong}, and G.~K. {Karagiannidis},
  ``Energy-efficient device discovery in {D2D} cellular networks for public
  safety scenario,'' \emph{IEEE Syst. J.}, pp. 1--4, 2019.

\bibitem{AD3}
Z.~{Kaleem}, M.~{Yousaf}, A.~{Qamar}, A.~{Ahmad}, T.~Q. {Duong}, W.~{Choi}, and
  A.~{Jamalipour}, ``{UAV}-empowered disaster-resilient edge architecture for
  delay-sensitive communication,'' \emph{IEEE Network}, pp. 1--9, 2019.

\bibitem{AD4}
M.~Z. {Anwar}, Z.~{Kaleem}, and A.~{Jamalipour}, ``Machine learning inspired
  sound-based amateur drone detection for public safety applications,''
  \emph{IEEE Trans. Veh. Technol.}, vol.~68, no.~3, pp. 2526--2534, Mar. 2019.

\bibitem{RP_170779}
3GPP, ``{RP}-170779: Study on enhanced support for aerial vehicles.''

\bibitem{Cache_Not}
J.~Erman, A.~Gerber, M.~Hajiaghayi, D.~Pei, S.~Sen, and O.~Spatscheck, ``To
  cache or not to cache: The {3G} case,'' \emph{IEEE Internet Comput.},
  vol.~15, no.~2, pp. 27--34, Mar. 2011.

\bibitem{D2D_Assisted_Caching}
Q.~{Li}, Y.~{Zhang}, A.~{Pandharipande}, X.~{Ge}, and J.~{Zhang},
  ``{D2D}-assisted caching on truncated {Z}ipf distribution,'' \emph{IEEE
  Access}, vol.~7, pp. 13\,411--13\,421, 2019.

\bibitem{D2D_caching}
D.~{Malak}, M.~{Al-Shalash}, and J.~G. {Andrews}, ``Spatially correlated
  content caching for device-to-device communications,'' \emph{IEEE Trans.
  Wireless Commun.}, vol.~17, no.~1, pp. 56--70, Jan. 2018.

\bibitem{Caching_NTier}
K.~Li, C.~Yang, Z.~Chen, and M.~Tao, ``Optimization and analysis of
  probabilistic caching in {$N$}-tier heterogeneous networks,'' \emph{IEEE
  Trans. Wireless Commun.}, vol.~17, no.~2, pp. 1283--1297, Feb. 2018.

\bibitem{Contract}
J.~{Li}, S.~{Chu}, F.~{Shu}, J.~{Wu}, and D.~N.~K. {Jayakody}, ``Contract-based
  small-cell caching for data disseminations in ultra-dense cellular
  networks,'' \emph{IEEE Trans. Mobile Comput.}, vol.~18, no.~5, pp.
  1042--1053, May 2019.

\bibitem{Distributed_Caching}
J.~Li, Y.~Chen, Z.~Lin, W.~Chen, B.~Vucetic, and L.~Hanzo, ``Distributed
  caching for data dissemination in the downlink of heterogeneous networks,''
  \emph{IEEE Trans. Commun.}, vol.~63, no.~10, pp. 3553--3568, Oct. 2015.

\bibitem{Mobile_Data_Caching}
K.~Poularakis, G.~Iosifidis, and L.~Tassiulas, ``Approximation algorithms for
  mobile data caching in small cell networks,'' \emph{IEEE Trans. Commun.},
  vol.~62, no.~10, pp. 3665--3677, Oct. 2014.

\bibitem{Pricing}
J.~{Li}, H.~{Chen}, Y.~{Chen}, Z.~{Lin}, B.~{Vucetic}, and L.~{Hanzo},
  ``Pricing and resource allocation via game theory for a small-cell video
  caching system,'' \emph{IEEE J. Sel. Areas Commun.}, vol.~34, no.~8, pp.
  2115--2129, Aug. 2016.

\bibitem{Cache_enabled_SCN}
E.~Ba{\c s}tu{\u g}, M.~Bennis, and M.~Debbah, ``Cache-enabled small cell
  networks: Modeling and tradeoffs,'' in \emph{EURASIP J. Wireless Commun.
  Netw.}, vol. 2015, no.~1, Aug. 2015, p.~41.

\bibitem{caching_sky}
M.~Chen, M.~Mozaffari, W.~Saad, C.~Yin, M.~Debbah, and C.~S. Hong, ``Caching in
  the sky: Proactive deployment of cache-enabled unmanned aerial vehicles for
  optimized quality-of-experience,'' \emph{IEEE J. Sel. Areas Commun.},
  vol.~35, no.~5, pp. 1046--1061, May 2017.

\bibitem{LTE_sky}
B.~V.~D. Bergh, A.~Chiumento, and S.~Pollin, ``{LTE} in the sky: Trading off
  propagation benefits with interference costs for aerial nodes,'' \emph{IEEE
  Commun. Mag.}, vol.~54, no.~5, pp. 44--50, May 2016.

\bibitem{UAV_Suburban}
A.~Al-Hourani and K.~Gomez, ``Modeling cellular-to-{UAV} path-loss for suburban
  environments,'' \emph{IEEE Wireless Commun. Lett.}, vol.~7, no.~1, pp.
  82--85, Feb. 2018.

\bibitem{Learn_Cache}
Z.~Chang, L.~Lei, Z.~Zhou, S.~Mao, and T.~Ristaniemi, ``Learn to cache: Machine
  learning for network edge caching in the big data era,'' \emph{IEEE Wireless
  Commun.}, vol.~25, no.~3, pp. 28--35, Jun. 2018.

\bibitem{Caching_UAV}
N.~Zhao, F.~Cheng, F.~R. Yu, J.~Tang, Y.~Chen, G.~Gui, and H.~Sari, ``Caching
  {UAV} assisted secure transmission in hyper-dense networks based on
  interference alignment,'' \emph{IEEE Trans. Commun.}, vol.~66, no.~5, pp.
  2281--2294, May 2018.

\bibitem{UAV_MIMO}
G.~Geraci, A.~G. Rodriguez, L.~G. Giordano, D.~L{\' o}pez-P{\' e}rez, and
  E.~Bj{\" o}rnson, ``Understanding {UAV} cellular communications: From
  existing networks to massive {MIMO},'' \emph{IEEE Access}, vol.~6, pp.
  67\,853--67\,865, 2018.

\bibitem{TR_36777}
3GPP, ``{TR} 36.777: Enhanced {LTE} support for aerial vehicles,'' Dec. 2017.

\bibitem{Tractable_Approach}
J.~G. Andrews, F.~Baccelli, and R.~K. Ganti, ``A tractable approach to coverage
  and rate in cellular networks,'' \emph{IEEE Trans. Commun.}, vol.~59, no.~11,
  pp. 3122--3134, Nov. 2011.

\bibitem{KTier_Downlink}
H.~S. Dhillon, R.~K. Ganti, F.~Baccelli, and J.~G. Andrews, ``Modeling and
  analysis of {K}-tier downlink heterogeneous cellular networks,'' \emph{IEEE
  J. Sel. Areas Commun.}, vol.~30, no.~3, pp. 550--560, Apr. 2012.

\bibitem{Optimal_geographic_caching}
B.~Blaszczyszyn and A.~Giovanidis, ``Optimal geographic caching in cellular
  networks,'' in \emph{Proc. IEEE ICC}, Jun. 2015, pp. 3358--3363.

\bibitem{Probabilistic_SC_Caching}
Y.~Chen, M.~Ding, J.~Li, Z.~Lin, G.~Mao, and L.~Hanzo, ``Probabilistic
  small-cell caching: Performance analysis and optimization,'' \emph{IEEE
  Trans. Veh. Technol.}, vol.~66, no.~5, pp. 4341--4354, May 2017.

\bibitem{Caching_Multicasting}
Y.~Cui and D.~Jiang, ``Analysis and optimization of caching and multicasting in
  large-scale cache-enabled heterogeneous wireless networks,'' \emph{IEEE
  Trans. Wireless Commun.}, vol.~16, no.~1, pp. 250--264, Jan. 2017.

\bibitem{LoS_NLoS}
M.~Ding, P.~Wang, D.~L{\' o}pez-P{\' e}rez, G.~Mao, and Z.~Lin, ``Performance
  impact of {LoS} and {NLoS} transmissions in dense cellular networks,''
  \emph{IEEE Trans. Wireless Commun.}, vol.~15, no.~3, pp. 2365--2380, Mar.
  2016.

\bibitem{Small_Cell_Caching}
J.~Li, Y.~Chen, M.~Ding, F.~Shu, B.~Vucetic, and X.~You, ``A small-cell caching
  system in mobile cellular networks with {LoS} and {NLoS} channels,''
  \emph{IEEE Access}, vol.~5, pp. 1296--1305, 2017.

\bibitem{Antenna_Heights}
M.~Ding and D.~L{\' o}pez-P{\' e}rez, ``Performance impact of base station
  antenna heights in dense cellular networks,'' \emph{IEEE Trans. Wireless
  Commun.}, vol.~16, no.~12, pp. 8147--8161, Dec. 2017.

\bibitem{Femtocell_BS}
I.~Ashraf, L.~T.~W. Ho, and H.~Claussen, ``Improving energy efficiency of
  femtocell base stations via user activity detection,'' in \emph{Proc. IEEE
  Wireless Commun. and Network. Conf.}, Apr. 2010, pp. 1--5.

\bibitem{TR_36828}
3GPP, ``{TR} 36.828: Further enhancements to {LTE} time division duplex ({TDD})
  for downlink-uplink {(DL-UL)} interference management and traffic
  adaptation,'' Jun. 2012.

\bibitem{Idle_Mode}
M.~Ding, D.~L{\' o}pez-P{\' e}rez, G.~Mao, and Z.~Lin, ``Performance impact of
  idle mode capability on dense small cell networks,'' \emph{IEEE Trans. Veh.
  Technol.}, vol.~66, no.~11, pp. 10\,446--10\,460, Nov. 2017.

\bibitem{Edge_Caching}
Q.~{Li}, W.~{Shi}, X.~{Ge}, and Z.~{Niu}, ``Cooperative edge caching in
  software-defined hyper-cellular networks,'' \emph{IEEE J. Sel. Areas
  Commun.}, vol.~35, no.~11, pp. 2596--2605, Nov. 2017.

\bibitem{limit}
M.~Ding, D.~L{\' o}pez-P{\' e}rez, G.~Mao, and Z.~Lin, ``Ultra-dense networks:
  Is there a limit to spatial spectrum reuse?'' in \emph{2018 IEEE
  International Conference on Communications (ICC)}, May 2018, pp. 1--6.

\end{thebibliography}

\end{document}